\documentclass[a4paper]{jpconf}
\usepackage{citesort}
\usepackage{graphicx}
\usepackage{amssymb}
\usepackage{amsmath}
\usepackage{amsfonts}
\usepackage{sidecap}
\usepackage{epstopdf}%\usepackage{color}
\usepackage{epsfig}
\usepackage{lineno}
\usepackage{xcolor}
\usepackage{soul} %to add deleted text JUANDRAFT
\newcommand{\doublefig}{0.49\textwidth} %0.8*8.5\columnwidth paper format
\newcommand{\imagesize}{0.6\textwidth} %0.8\columnwidth paper format

\newcommand{\ket}[1]{|#1\rangle}

\def\cn2{|c_n|^2}

\newcommand{\K}{K$^+$\,}
\newcommand{\Kforty}{$^{40}$K\,}
\newcommand{\ii}{\textrm{i}}

\begin{document}
%\nocite{eam}

%\newcommand{\C}{$^\circ$C}
%\newcommand{\Cmath}{^\circ\text{C}}
\newcommand{\juancomment}[1]{\textcolor{green}{\bf Juan Comment: #1}}  %writes in bf with Juan in front of the text
\newcommand{\juan}[1]{{#1}} %uncomment this command and comment the previous one to eliminate the Juan and bf.
\newcommand{\juandeleted}[1]{\st{#1}}
\newcommand{\comment}[1]{{ \bf Comment: #1}}  %writes in bf with Juan in front of the text
\newcommand{\fracc}[2]{\frac{\displaystyle #1}{\displaystyle #2}} %uncomment this command and comment the previous one to eliminate the Juan and bf.
\title{A semiclassical model for charge transfer along ion chains in silicates}%
\author{Juan F R Archilla$^{1,*}$, J\={a}nis Baj\={a}rs$^{2}$, Yusuke Doi$^3$ and Masayuki Kimura$^4$}% J\={a}nis Baj\={a}rs

\address{$^1$ Group of Nonlinear Physics, Universidad de Sevilla, ETSII, Avda Reina Mercedes s/n, 41012-Sevilla, Spain}
\address{$^2$ Faculty of Physics, Mathematics and Optometry, University of Latvia, Jelgavas Street 3, Riga, LV-1004, Latvia}
\address{$^3$ Division of Mechanical Engineering, Graduate School of Engineering, Osaka University, 2-1 Yamadaoka, Suita, Osaka 565-0871, Japan}
\address{$^4$ Department of Electrical and Electronic Engineering,  Faculty of Science and Engineering, Setsunan University, 17-8 Ikeda-Nakamachi, Neyagawa, Osaka 572-8508, Japan}

\ead{$^*$Corresponding author: archilla@us.es}

 \begin{abstract} %200 words. There are 259.
 It has been observed in fossil tracks and experiments in the layered silicate mica muscovite the transport of charge through the cation layers sandwiched between the layers of tetrahedra-octahedra-tetrahedra.  A classical model for the propagation of anharmonic vibrations along the cation chains has been proposed based on first principles and empirical functions. In that model, several propagating entities have been found as kinks or crowdions and breathers, both with or without wings, the latter for specific velocities and energies. Crowdions are equivalent to moving interstitials and transport electric charge if the moving particle is an ion, but they also imply the movement of mass, which was not observed in the experiments. Breathers, being just vibrational entities, do not transport charge. In this work, we present a semiclassical model obtained by adding a quantum particle, electron or hole to the previous model. We present the construction of the model based on the physics of the system. In particular, the strongly nonlinear vibronic interaction between the nuclei and the extra electron or hole is essential to explain the localized charge transport, which is not compatible with the adiabatic approximation. The formation of vibrational localized charge carriers breaks the lattice symmetry group in a similar fashion to the Jahn-Teller Effect, providing a new stable dynamical state.
 We study the properties and the coherence of the model through numerical simulations from initial conditions obtained by tail analysis and other means. We observe that although the charge spreads from an initial localization in a lattice at equilibrium, it can be confined basically to a single particle when coupled to a chaotic quasiperiodic breather. This is coherent with the observation that experiments imply that a population of charge is formed due to the decay of potassium unstable isotopes.
\end{abstract}
%\begin{keyword}
%nonlinear waves\sep kinks\sep crowdions\sep breathers\sep ILMs\sep nanopterons\sep charge transport
%nonlinear waves \sep polarobreathers \sep charge transport \sep silicates \sep mica muscovite
%\PACS  63.20.Pw %Localized modes
% \sep 63.20.Ry  % Anharmonic lattice modes
%\sep 05.45.-a	%Nonlinear dynamics and chaos
%02.70.-c %Computational techniques in mathematical methods in physics
%\sep 64.70.kp	%Ionic crystals
% \sep 63.22.Np %layered systems
%\maketitle
%\end{keyword}

\section{Introduction}
\label{sec:introduction}
Tracks in mica muscovite were observed due to the special properties of the material. It can be exfoliated into very thin sheets, which are also semitransparent and therefore allow for direct observation. Some of those dark tracks were due to positively charged particles such as positrons, protons, or antipions\,\cite{pricewalker62,russell67a,russell67b}. Many other tracks were along the potassium hexagonal lattice layer and were therefore attributed to quasi-one-dimensional lattice excitations called quodons\,\cite{russell-collins95a,russellnaturereview2022}. They were experimentally observed in an experiment where alpha particles were sent onto a monocrystal side, and subsequently, it was observed the ejection of an atom from the other side along the lattice close-packed directions\,\cite{russell-experiment2007}. Fossil tracks were most probably produced by the recoil of potassium ions after beta decay, which is 99\% electron emission,  leaving behind a positive charge. Therefore, it was deduced that quodons could transport electric charge\,\cite{russell-tracks-quodons2015article,archillaLoM2016}, which opened the possibility of the experimental measurement of electric current. This was achieved by bombarding a mica monocrystal with alpha particles which were expected to produce a large number of nonlinear excitations and measuring the current in the absence of an electric field, a phenomenon called {\em hyperconductivity}. There was an initial peak of current that after some seconds would diminish to the current transported by the flux of alpha particles\,\cite{russell-archilla2017}. It was interpreted as the nonlinear excitations trapping the accumulated electric charge left behind from electron beta emission. When this reservoir was exhausted, the only charge available was the one provided by the alpha particles. More experiments were done with other silicates\,\cite{russell-archilla2019} and other materials as it was developed a test to distinguish hyperconductivity and to separate quodon currents from Ohmic currents in semiconductors or conductors\,\cite{russell-archilla2021rrl,russell-archilla2022ltp}. The authors recommend a recent review on the subject\,\cite{russell-archilla2021springer}.

A classical model for a cation chain in muscovite was developed based on first principles and empirical potentials. Interestingly, it was found the existence of crowdions or lattice kinks with energies of 26\,eV, which is below the energy provided by the nucleus recoil after beta emission and above the energy to eject an atom, therefore coherent with the mica ejection experiment\,\cite{archilla-kosevich-springer2015article,archilla-kosevich-pre2015}. There were also found kinks with other energies but with wings, also called nanopterons\,\cite{remoissenet1999,archilla-zolotaryuk2018}. Interestingly, a crowdion is a moving interstitial, which within an ionic crystal implies the movement of electric charge, making  it a candidate for hyperconductivity. There are, however, observations of primary fossil tracks, which scatter in many other fainter ones, which should have much smaller energies, implying that less energetic nonlinear excitations as breathers were also of interest.

Breathers are nonlinear localized solutions with also a vibration\,\cite{flach2008}. They are well described mathematically\,\cite{mackayaubry94}, and there are methods to construct numerically exact ones\,\cite{marinaubry96}. They can also appear in systems with long-range interaction\,\cite{archilla-bentchain2001} as alpha-helix proteins\,\cite{archilla-alpha-helix2022}.
Breathers, also known as {\em intrinsic localized modes} (ILMs), were found in 3D in Si\,\cite{voulgarakis2004}. The theory of ILMs in 3D molecular dynamics was developed in Ref.\,\cite{hizhnyakov-scripta2014}, where it was found that they appear after X-ray recoil in molecular dynamics of ionic crystals but also in metals, such as Ni, Nb, and Fe. It was demonstrated that ILM frequencies can be within gaps in the phonon band or above it. The ILMs were highly mobile with energy of the order of 1\,eV.  ILMs were later found in other bcc metals, such as V and W\,\cite{murzaev-bcc-2015}, in fcc crystals, such as Cu\,\cite{hizhnyakov2015,hizhnyakovLoM2016}, and in hcp Be\,\cite{bachurinagmoving2018}. ILMs were also found in covalent crystals with the diamond structure, such as the insulator C, and the semiconductors Si and Ge\,\cite{hizhnyakov2015,hizhnyakovLoM2016}. They were also constructed in graphene both with classical\,\cite{hizhnyakovPLA2016},  and  ab initio molecular dynamics\,\cite{lobzenko2016}. To summarize, it was demonstrated that breathers or ILMs can appear in many materials with different electric behavior and with many different crystal structures.

The spectral theory of breathers in the moving frame was developed in Ref.\,\cite{archilla-osaka19}. In the same work, it was applied to the muscovite model and traveling exact breathers with small energy were found.  A related phenomenological model for muscovite was also used with the property that it was very easy to obtain traveling breathers in two dimensions\,\cite{bajars-springer-article2015}. The theory of exact breathers in the moving frame was extended to two and more dimensions using that model as an  example\,\cite{bajars-archilla2022A}. The theory  was also extended to a variation of the latter model with the addition of a quantum particle. Numerical methods were developed to deal with the differences in the time scale of the charge and the atoms that were able to conserve the charge probability\,\cite{bajars-archilla2022B}. Their spectral properties were deduced and described in Ref.\,\cite{archilla-bajars2023}.

In this paper, we construct a semiclassical model for the specific model based on potassium chains in mica muscovite constructed from first principles and empirical potentials in Ref.\,\cite{archilla-kosevich-pre2015}, by adding a quantum particle. In many aspects, the model is similar to the phenomenological model used in Ref.\,\cite{bajars-archilla2022B,archilla-bajars2023}, but it is more complicated as the diagonal terms of the quantum charge Hamiltonian are not constant but correspond to the interaction with the other electric charges in the crystal.

The vibronic interaction between the nuclei and the extra electron or hole is strongly nonlinear with the consequence that the tunneling or an electron between nearest--neighbor ions is only probable when the nuclei are close enough and the transition probability increases strongly as they approach. The consequence is the formation of quodons, dynamical states that transport electric charge in a localized manner, breaking the lattice discrete translation invariance as happens with the spontaneous symmetry breaking in the Jahn-Teller or pseudo Jahn-Teller effect\cite{bersuker2006,Bersuker2017}. Certainly, the space of quodons has to keep the lattice translation invariance but given the low probability for the formation of quodons, their population at any given time should also break the translational invariance.

The paper is structured as follows: after the introduction, the model is described in Section\,\ref{sec:model}, including the hole or electron potential; the Hamiltonian and dynamical equations are then obtained in Section\,\ref{sec:hamdynequations}, which are linearized in Section\,\ref{sec:linear}.  Section\,\ref{sec:simulationtests} presents the results of numerical integration for different initial variables, such
as localized traveling and stationary trial solutions, and localized charge in a system at equilibrium; also it is observed a breather rebounding in a charge and a kink with an extra charge, and finally, a charge trapped by a chaotic quasiperiodic breather. The main part of the article finishes with the conclusions. There are also two appendixes: Appendix A describes the transformation of the semiclassical system into a real canonical Hamiltonian one, and Appendix B describes methods for numerical integration that preserve the charge probability
at each integration step.

\section{Model}
\label{sec:model}
We propose a tight-binding model for a positive charge, a hole, or an electron, that we will call a charge with $Q=1$ for the hole and $Q=-1$ for the electron. The model is an extension of the model already used for muscovite by Archilla et al.\,\cite{archilla-kosevich-pre2015,archilla-kosevich-springer2015article,archilla-zolotaryuk2018,archilla2019}. The ket $\ket{n}$ represents the state in which the extra charge is located at the lattice position $n$, being $\langle n|$ the corresponding bra or adjoint operator, with $\langle m|n\rangle =\delta_{m,n}$. An extra charge state is given by $|\phi(t)\rangle=\sum_n c_n(t)|n\rangle$, where $c_n(t)$ represent the time-dependent possibility that the extra charge is located at site $n$. We will omit in what follows the explicit time dependence of $|\phi\rangle$ and $c_n$ when convenient. The adjoint operator to $|\phi \rangle$ is
$\langle\phi|=\sum_n c^*_n\langle n|$, where $c_n^*$ is the complex conjugate of $c_n$ and $|c_n^2|=c_n^*c_n$ is the probability of finding the extra charge in the site $n$ in the state $|\phi\rangle$. The total probability is one as there is an electron or hole in the system, i.e.:
\begin{linenomath}\begin{equation}\label{eq:charge}
\sum_{n=1}^{N}|c_n|^2=1.
\end{equation}\end{linenomath}

The cations \K are subjected both to an on-site potential $U(u_n)$ which represents the interaction of \K with the ions of the surrounding lattice except for the \K ions in the same row, which are explicitly included as the interatomic interaction $V(u_n-u_{n-1})=V_C(u_n-u_{n-1})+V_Z(u_n-u_{n-1})$, where $V_C$ is the Coulomb repulsion; and $V_Z$, a Yukawa type potential is a simplification of the ZBL potential\,\cite{ziegler2008}, and corresponds to the repulsion between nuclei screened by the electron cloud. The variables $u_n$ represent the $n$-th ion separation from the equilibrium position. We will use scaled units convenient for the modeled system: $u_L=a=5.19$\,\AA\, is the equilibrium distance between potassium ions; $u_E$ is the Coulomb energy corresponding to two units of charge $e$ at distance $u_L$, i.e., $u_E=k_e e^2/a \simeq 2.77$\,eV; the unit of mass is the mass of a potassium atom $u_M=m_\text{K}=39.1$\,amu; and the unit of time becomes a derived quantity: $u_T=(u_M u_L^2/uE)^{1/2} = (m_K a^3/k_e e^2 )^{1/2}\simeq 0.2$\, ps. The unit of angular frequency is equivalent
to 5\,Trad/s or $u_f=1/(2\pi u_T )\simeq 0.8$\,THz$\simeq 26.7$\,cm$^{-1}$.

In the scaled units, the interaction potential energies become:
\begin{eqnarray}
V &=&  V_C+V_Z\label{eq:Vsum}\\
V_C(u_n-u_{n-1})&=&\frac{1}{1+u_n-u_{n-1}}\label{eq:VCoulomb} \\
V_Z(u_n-u_{n-1})&=&\frac{B}{1+u_n-u_{n-1}}\exp(-\frac{1+u_n-u_{n-1}}{\rho})=\nonumber\\
&\mbox{}&\frac{C}{1+u_n-u_{n-1}}\exp(-\beta(u_n-u_{n-1}))\, ,
\label{eq:Vzbl}
\end{eqnarray}
with  $B=184.1$  %$B=184.0739$
and $\rho=0.05690$ %$\rho=0.05689672128208$,
$\beta=1/\rho=17.58$  %$\beta=17.575705198235326$
and $C=B\exp(-\beta)=4.285\times 10^{-6}$. % $C=B\exp(-\beta)=4.285094764213043\times 10^{-6}$.

The on-site potential is obtained with the use of empirical potentials and electrostatic potentials\,\cite{gedeon2002}, considering the interaction with the ions in the layers above, the first one composed of O$^{-2}$, and the second of a mixed species between Si and Al with a positive charge $+3.1$. The resulting potential is a Fourier series that can be truncated at the fourth term:
\begin{eqnarray}
U=\sum_{m=0}^4 v_m\cos(2\pi x), \quad \textrm{with} \quad \\
\{v_m\}=\{2.4474,-3.3490, 1.0997,-0.2302, 0.0321\}\,.
\label{eq:Uonsite}
\end{eqnarray}
The on-site potential $U$ is represented among other potentials in Fig.\,\ref{fig:Uhole}. It has a potential barrier of 20\,eV, as obtained with molecular dynamics\,\cite{CollinsAM92}, and a small amplitude frequency of 110\,cm$^{-1}$, as observed experimentally\,\cite{diaz2000}. It is soft for $x\lesssim 0.3a$, and hard for larger distances as corresponds to a bounded system.

 Both the interaction and the on-site potentials are described in detail in Refs.~\cite{archilla-kosevich-pre2015,archilla-osaka19}. In the following, we consider for the first time some of the consequences of adding an extra unit of charge as a hole or an electron.
\subsection{Hole or electron potential}
\label{subsec:holepotential}
Localized energy transport in muscovite has been observed experimentally\,\cite{russell-experiment2007}, and it has been deduced that dark tracks are produced by positive charge\,\cite{russell-tracks-quodons2015article,archillaLoM2016}. Subsequently, charge transport has been observed experimentally in muscovite\,\cite{russell-archilla2017} and other silicates\,\cite{russell-archilla2019,russell-archilla2021springer,russell-archilla2021rrl,russell-archilla2022ltp}. Different tracks suggest different types of carriers\,\cite{Russell2018}. In this article, we attempt to model the transport of positive or negative charge attached to the \K\,ions and coupled with the lattice.

An extra charge will appear due to $\beta^-$ decay of \Kforty\,\cite{radionuclides2012}, for which the nucleus transforms in $^{40}$Ca with an extra proton. The ion becomes Ca$^{++}$, with the extra hole migrating to the neighboring ion. The far less frequent $\beta^-$ decay will transform \Kforty into $^{40}$Ar, and the ion will become Ar$^{0}$, with an electron that can migrate to neighboring \K transforming them into neutral K \cite{archillaLoM2016}.
\begin{figure}[t]
\begin{center}
\includegraphics[width=0.9\textwidth]{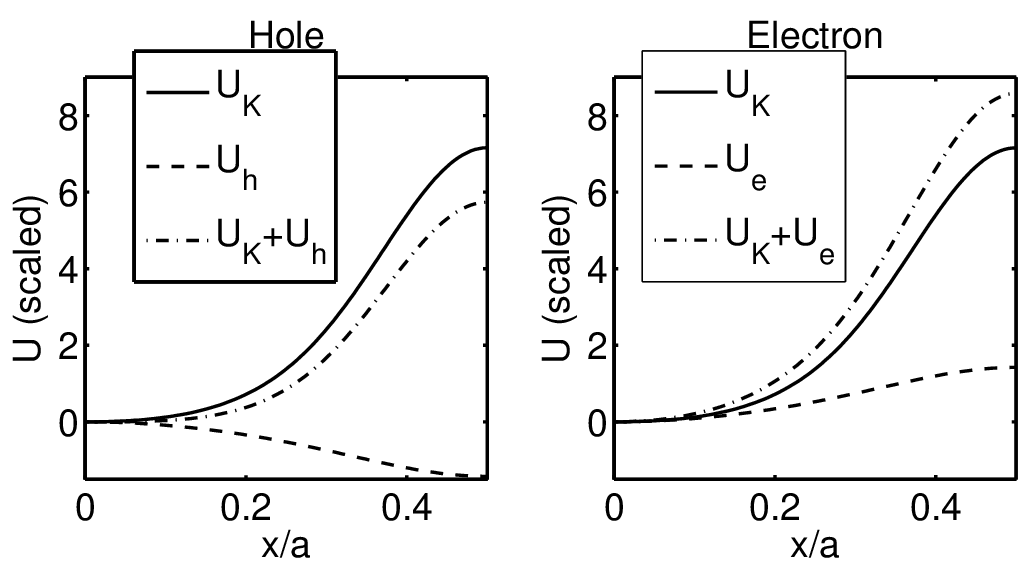}
\end{center}
\caption{({\bf Left}) Substrate potentials $U_K$ experienced by a potassium ion K$^+$; $U_h$, the electric potential experienced by a positive hole h$^+$;  and $U_k+U_h$, is the potential experienced by the double cation K$^{++}$. $U_K$ is a sum of electrical and Buckingham terms\,\cite{gedeon2002}.
({\bf Right}) Also, $U_K$ is represented with $U_e$, the potential experienced by an extra electron  e$^-$,  and $U_K+U_e$, the potential experienced by the neutral atom K$^{0}$. The latter is therefore due only to the Buckingham potentials in Ref.\,\cite{gedeon2002}.
The unit of energy is $u_E\simeq 2.77$\,eV.
 }%end caption
\label{fig:Uhole}
\end{figure}

An extra charge in site $n$ will experience the electrostatic interaction with the surrounding ions, given by $U_Q(u_n)$,  and also with the other \K in the same row, given also by $V_C(u_n-u_{n-1})$ in \eqref{eq:VCoulomb}.  $U_Q$  is different from $U$ in \eqref{eq:Uonsite}, because the short-range interaction is already taken into account, and therefore only the extra electrostatic interaction of the extra charge has to be considered~\cite{gedeon2002}. Considering the interaction with the oxygen ions at the immediate layers above and below and with the Si-Al mixed species in the following layer with charge $+3.1$ leads to a potential described also by a Fourier series truncated at the fourth harmonic. We denote it as $U_Q$, being $U_h=U_Q$ for a hole and $U_e=-U_h$ for an extra electron. They are given by:
\begin{eqnarray}
U_h&=&\sum_{m=0}^4 h_m\cos(2\pi x), \quad \textrm{with}\nonumber\\
\{h_m\}&=&\{-0.6160, 0.6941, -0.0930, 0.0167, -0.0018\}\,,\\
U_Q&=&QU_h \, .  \nonumber
%\{h_m\}&=&\{-0.615939937753237, 0.694075384744464, -0.093049835093594, \nonumber\\
%&\mbox{}&0.016699537003682, -0.001785148901316\}\,.
\label{eq:Uhonsite}
\end{eqnarray}
Figure~\ref{fig:Uhole}-Left  shows the different potentials. Note that $U_h$ has a maximum at the equilibrium position for a hole ($Q=+1$) because it is energetically favorable to become closer to one of the negative oxygen ions. On the contrary, for an electron $Q=-1$, the potential energy $U_e$ has a minimum.

The potential $U_h$ has a maximum at $x=0$ and a minimum between at $x=0.5$ with a potential well of $\simeq -3.74$\,eV, which diminishes the height of the potential barrier. The potential $U_e=-U_h$ has the opposite properties.

 The potential  $V_C(u_n-u_{n-1})$ is equal to \eqref{eq:VCoulomb}, as is the Coulomb repulsion between an extra charge in site $n$ and the nearest \K ions. In this case, the interaction with the extra charge with the \K ions in the same row does not include a Yukawa or ZBL potential because there is not an extra nuclei.

\begin{figure}[b]
\begin{center}
\includegraphics[width=\imagesize]{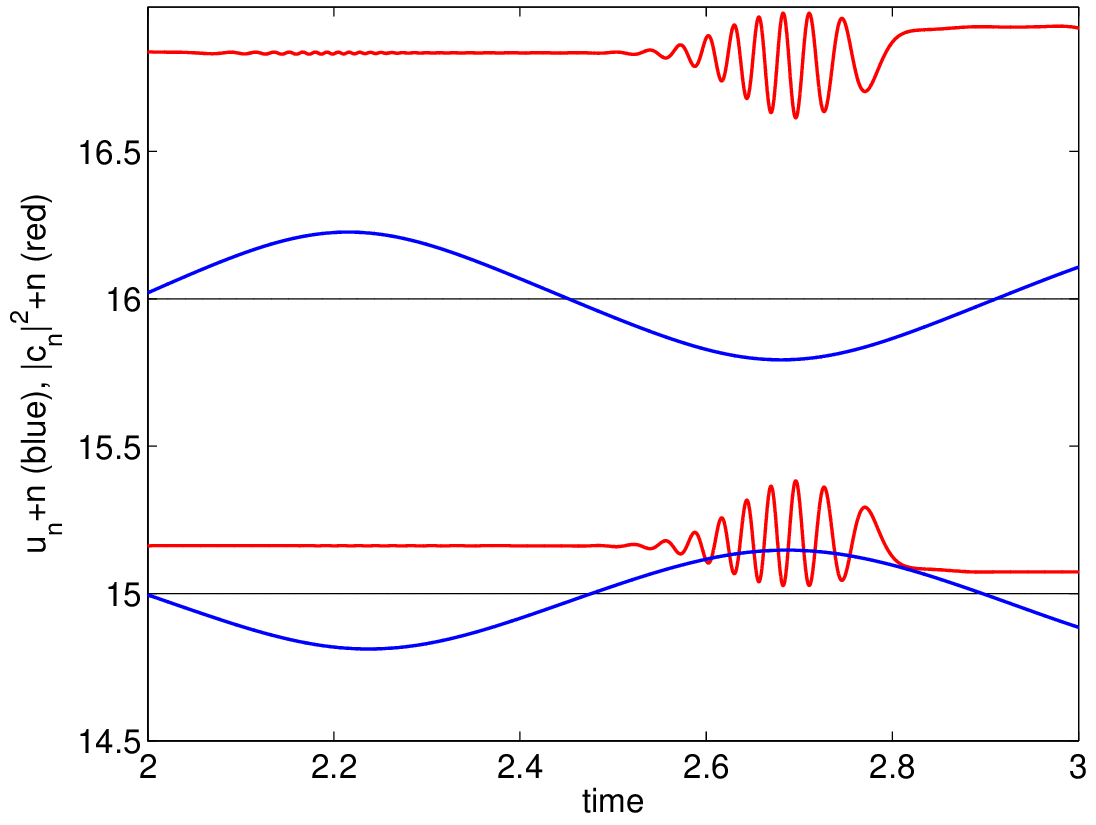}
\end{center}
\caption{Displacements (blue lines) and charge probability (red lines) for a system after an initial compression of particles 15 and 16 to a distance of $0.4$. It brings about the oscillations of the particles in the antiphase for some time. The interchange of probability when the ions approach can be seen, and also the high frequency of the charge transfer. }%end caption
\label{fig_approaching}
\end{figure}

Therefore the Hamiltonian operator for the extra charge will be
\begin{equation}
\hat H_Q=\sum_n E_n|n\rangle \langle n| -J_{n,n+1}|n\rangle \langle n+1|-J_{n,n-1}|n\rangle\langle n-1| \, .
\end{equation}
The transfer integrals $J_{n,n-1}=J_{n-1,n}$ are related to the probability of a transition from the state $|n-1\rangle$ to the state $|n\rangle$ and vice versa.  The term $E_n$ can be obtained easily   as the  expected value of the charge Hamiltonian  $\langle n|\hat H_Q|n\rangle$ when the nondiagonal terms $J_{n,n-1}$ are $zero $. It will be composed of the classical energy of the extra charge in the site $n$, that is, the electrostatic interaction with the lattice, plus the electrostatic interaction with the nearest \K ions. Then:
 \begin{equation}
E_n=QU_h(u_n)+ \frac{Q}{1+u_n-u_{n-1}}+\frac{Q}{1+u_{n+1}-u_n}-2Q+E_0\, .   %V_C(u_n-u_{n-1})+V_C(u_{n+1}-u_n) \, .
\label{eq:En}
\end{equation}
We have subtracted the hole electrostatic energy at equilibrium $-2Q$ and added a reference value $E_0$, so as $E_n=E_0$ at the equilibrium distance. The value of $E_0$ has no physical consequences, and it will be generally taken as zero, however, some other values may be more convenient than others for numerical integration and to obtain periodic solutions\,\cite{archilla-bajars2023}.

The terms $J_{n-1,n}$ have to be negligible for the equilibrium distance and become large only at the distance where the two electronic clouds of the \K ions interact. Therefore, a reasonable assumption is that they are exponentials with a decay rate similar to the ZBL repulsion between nuclei. That is,
\begin{equation}
J_{n-1,n}=J_{n,n-1}=J_0\exp(-\alpha(1+u_n-u_{n-1}))=I_0\exp(-\alpha(u_n-u_{n-1}))\, .
\label{eq:Jexp}
\end{equation}
A first guess of $\alpha$ is $\alpha=\beta$ as both terms are the consequence of the overlapping of the electron shells. In principle, $I_0$ is the same for a  hole or an electron, but this assumption might be revised.

Note that $I_0$ is the only parameter for which we do not have an approximate value at the moment. We expect to deduce it from the band structure and experimental mobilities in muscovite, but at this stage, it will be taken as an adjustable parameter. Also, the value of $\alpha$ initially equal to $\beta$ might have to be reconsidered.

Figure\,\ref{fig_approaching}  shows  the interchange of probability when the particles approach as a consequence of the functional form of the transfer integrals.

\section{Hamiltonian and dynamical equations}
\label{sec:hamdynequations}
We obtain the dynamical equations for the charge amplitudes from the Schr\"odinger equation $\ii \hbar \partial/\partial t |\phi\rangle=\hat H_Q\ket{\phi}$ collecting together the coefficients of the basis states $|n\rangle$. The Planck constant in scaled units will be denoted $\tau=\hbar/u_E/u_T$, where %$u_E=4.439306358381501\times10^{-19}$\,J$\simeq$  2.7746\,eV and $u_t=0.198476481628887$\,ps
$u_E$ and $u_T$ are the scaled units of energy and time, therefore $\tau=0.0011968$. %$\tau=0.001196798883201$.
Then:
\begin{equation}
\ii\tau\dot c_n=\left[QU_h(u_n)+\frac{Q}{1+u_n-u_{n-1}}+\frac{Q}{1+u_{n+1}-u_n}-2Q+E_0\right]c_n-\left[J_{n,n-1}c_{n-1}+J_{n,n+1}c_{n+1}\right]\, .
\label{eq:dynam1}
\end{equation}

The equations of movement for the variables $u_n$ are obtained from the Hamiltonian as $\dot p_n=-\partial H_{tot}/\partial u_n$  and $\dot u_n=\partial H_{tot}/\partial p_n=p_n$. Where $H_{tot}$ is the Hamiltonian obtained as

\begin{equation}
H_{tot}=H_{lat}+\langle \phi|H_Q|\phi\rangle.
\label{eq:Ham}
\end{equation}

The first component of the Hamiltonian is the lattice classical Hamiltonian:

\begin{eqnarray}
H_{lat}&={\displaystyle \sum_n \fracc{1}{2}p_n^2+U(u_n)+V(u_{n}-u_{n-1})}\, ,
 \label{eq:Hlat}
 \end{eqnarray}
 with $V$  and $U$ given in \eqref{eq:Vsum} and \eqref{eq:Uonsite}. The second component of the Hamiltonian is the expected value of the charge Hamiltonian in a generic state $\ket{\phi}=\sum_n c_n |n\rangle $, which is given by:
 \begin{eqnarray}
H_Q=\langle\phi|\hat H_Q|\phi \rangle &=&\sum_n E_n c_n^* c_n-\left[J_{n,n+1}c_n^* c_{n+1}+J_{n,n-1}c_n^* c_{n-1}\right]\, ,
\label{eq:H_h}
%H_Q &=& \sum_n E_n c_n^* c_n-J_{n,n+1}\left[c_n^* c_{n+1}+c_n c_{n+1}^*\right]\, \quad \mathrm{or}\nonumber\\
%H_Q &=& \sum_n E_n c_n^* c_n-2J_{n,n+1}\R \left[c_n^* c_{n+1}\right]\, \quad \mathrm{or}\nonumber\\
%H_Q &=& \sum_n E_n (a_n^2+b_n^2)-2J_{n,n+1}\left[a_n a_{n+1}+b_n b_{n+1}\right]\,,\\
\end{eqnarray}
with  $E_n$ from  \eqref{eq:En} and  $J_{n,n+1}$ from \eqref{eq:Jexp}.
%being the coefficients of $c_n$ in Eq.~\ref{eq:dynam1} and depend on $u_n,u_{n+1}$ and $u_{n-1}$.

%To obtain the dynamical equations we need the derivatives of $H_Q$; which are given by:
%with respect to $u_n$ of the previous Hamiltonian, it has to be taken into account $E_{n+1}$ and $E_{n-1}$
%\begin{eqnarray}
%\dpartial{H_Q}{u_n} &=\dpartial{E_n}{u_n}c_n^*c_n+\dpartial{E_{n+1}}{u_n}c_{n+1}^*c_{n+1} +\dpartial{E_{n-1}}{u_n}c_{n-1}^*c_{n-1}\nonumber\\
%- &\mbox{} \dpartial{J_{n+1,n}}{u_n}\left[ c_n^* c_{n+1}+c_{n+1}^*c_n \right]-\dpartial{J_{n,n-1}}{u_n}\left[ c_n^* c_{n-1}+c_{n-1}^*c_n\right]
%\end{eqnarray}
%where
%\begin{eqnarray}
%\dpartial{E_n}{u_n}=QU'_h(u_n)-QV_C'(u_{n+1}-u_{n})+QV_C'(u_n-u_{n-1}) \\
%\dpartial{E_{n+1}}{u_n}=-QV_C'(u_{n+1}-u_{n})=+\frac{Q}{(1+u_{n+1}-u_{n})^2}\\
%\dpartial{E_{n-1}}{u_n}=QV_C'(u_n-u_{n-1})=-\frac{Q}{(1+u_n-u_{n-1})^2}\\
%\dpartial{J_{n+1,n}}{u_n}=\alpha I_0\exp(-\alpha(u_{n+1}-u_n))\\
%\dpartial{J_{n,n-1}}{u_n}=-\alpha I_0\exp(-\alpha(u_{n}-u_{n-1}))\\
%\end{eqnarray}
\subsection{Final equations}
\label{subsec:finalequations}
Calculating the derivatives of $E_n$ and $J_{n,n+1}$, and collecting together the different terms, we obtain the dynamical equations for $u_n$:
\begin{eqnarray}
\ddot u_n=-U'(u_n)-QU'_h(u_n)|c_n|^2\nonumber\\
 -\frac{1}{(1+u_{n+1}-u_n)^2}\left[ 1+C\exp(-\beta(u_{n+1}-u_n))+Q|c_n|^2+Q|c_{n+1}|^2   \right]\nonumber\\
+\frac{1}{(1+u_{n}-u_{n-1})^2}\left[ 1+C\exp(-\beta(u_{n}-u_{n-1}))+Q|c_n|^2+Q|c_{n-1}|^2   \right]\nonumber\\
+\frac{C\beta\exp(-\beta(u_n-u_{n-1}))}{1+u_n-u_{n-1}}-\frac{C\beta\exp(-\beta(u_{n+1}-u_n))}{1+u_{n+1}-u_{n}}\nonumber \\
+\alpha I_0\exp(-\alpha(u_{n+1}-u_n))(c_{n+1}^*c_n+c_{n}^*c_{n+1})\nonumber\\
-\alpha I_0\exp(-\alpha(u_{n}-u_{n-1}))(c_{n}^*c_{n-1}+c_{n-1}^*c_n)\,.
\label{eq:unddot}
\end{eqnarray}
Note that the last two lines are real. %and can also be written as:

The dynamical equations for the extra charge are given by:
\begin{eqnarray}
\ii\tau\dot c_n=\left[QU_h(u_n)+\frac{Q}{1+u_n-u_{n-1}}+\frac{Q}{1+u_{n+1}-u_n}-2Q+E_0\right]c_n\nonumber\\
-I_0\exp(-\alpha(u_n-u_{n-1}))c_{n-1}- I_0\exp(-\alpha (u_{n+1}-u_{n})) c_{n+1}\, .
\label{eq:cndot}
\end{eqnarray}

These equations can be written in real form and also as canonical Hamiltonian equations as explained in \,\ref{sec:hamiltonianODE}.

\section{Linearization }
\label{sec:linear}
We expand the terms in the dynamical equations\,\eqref{eq:unddot}-\eqref{eq:cndot}, using $1/(1+x)\simeq 1-x+x^2$ and $1/(1+x)^2\simeq 1-2x+3x^3$, and  we can neglect  the ZBL potential because of small displacements $u_n$,  $C~\simeq 10^{-6}$ and $\beta C\simeq 10^{-4}$. We obtain the linearized dynamical equations:
\begin{eqnarray}
\ddot u_n&=&-\omega_0^2 u_n+2(u_{n+1}+u_{n-1}-2 u_n)\, ,\label{eq:unddotlinear}\\
\ii\tau\dot c_n&=&E_0 c_n-I_0[c_{n+1}+c_{n-1}]\, \label{eq:cndotlinear}.
\end{eqnarray}
The frequency of the lattice homogeneous oscillations $\omega_0$ is given by
\begin{equation}
\omega_0=(-\sum_{m=1}^4 (2\pi m)^2 v_m)^{1/2}\simeq 4.4800\, .%4.480047712921744\simeq 4.4800 \,  \\
\end{equation}

 A value of $E_0\neq 0$ implies only a shift in the $c_n$ frequencies of $E_0/\tau$, which can be convenient for integration purposes but has no physical consequences as always the products $c_nc_m^*$ that appear in the dynamical equations are invariant with respect to a global frequency shift\,\cite{archilla-bajars2023}.  Note that the variables $u_n$ and $c_n$ become decoupled at the linear limit.

The dispersion relations are independent and are given by
\begin{eqnarray}
\omega^2&=&\omega_0^2+4c_s^2\sin^2(q/2)\,,\quad \text{for the variables}\, u_n\,;\label{eq:DRu}\\
\omega&=&\frac{E_0}{\tau}-\frac{2I_0}{\tau}\, \cos(q)\,,\quad\text{for the variables}\, c_n\,.\label{eq:DRc}
\end{eqnarray}

The constant $c_s = \sqrt{2}$ in \eqref{eq:DRu}  is the sound velocity in the lattice system without on-site potential and it is written as a symbol for comparison with other scalings. The second equation\,\eqref{eq:DRc} multiplied by the scaled Planck constant $\tau$  provides the charge energy\,\cite{griffiths2018}, i.e.:
\begin{equation}
H_Q=\tau\omega=E_0-2 I_0\, \cos(q)\,.\label{eq:Hclinear}
\end{equation}

\begin{figure}[b]
\includegraphics[width=0.49\textwidth]{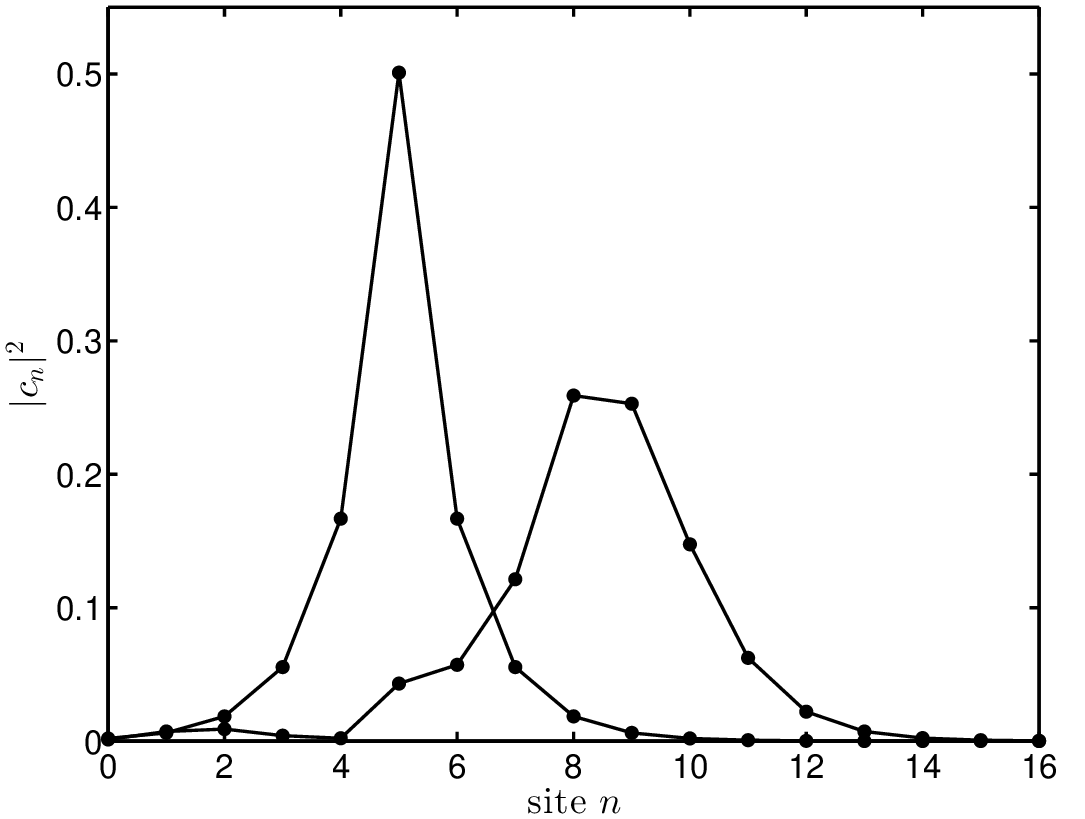}
\includegraphics[width=0.49\textwidth]{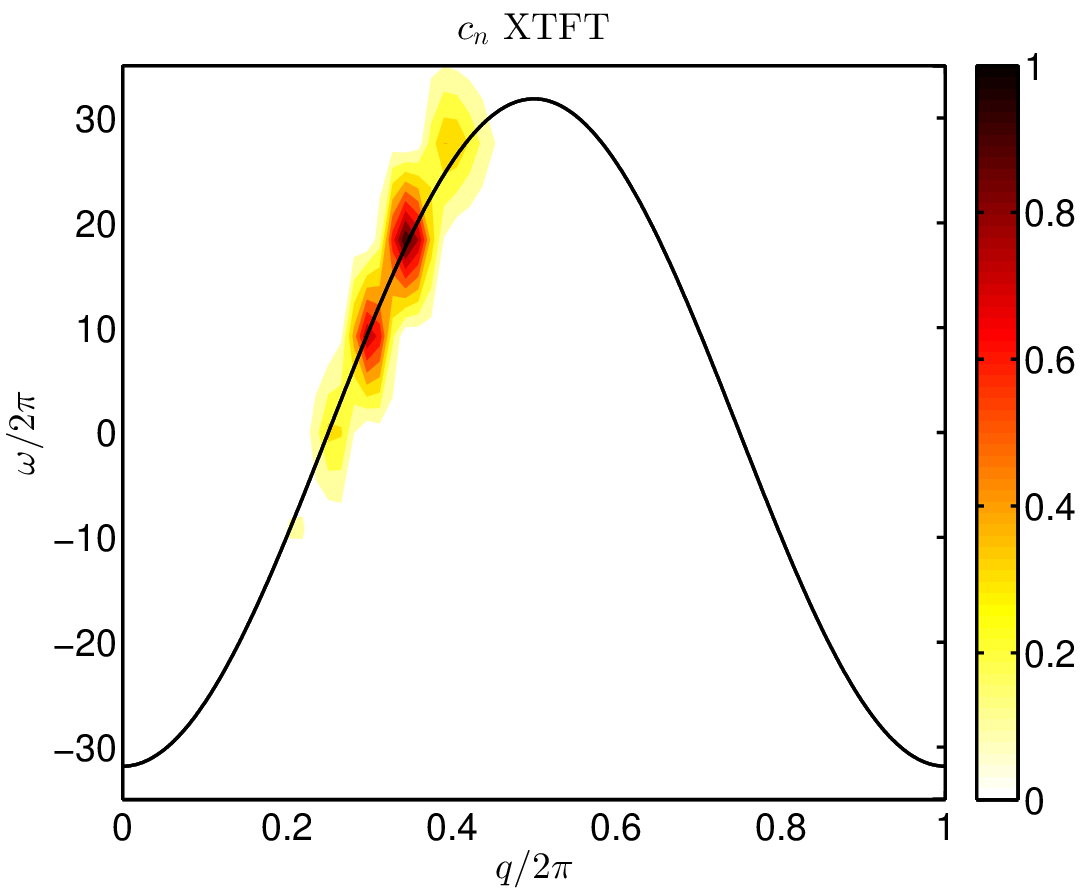}%{Figures/fig_localizedFFT.eps}
\caption{{\bf (Left)} Localized wave obtained with initial conditions $u=0$, $p=0$, $c_n=A\exp(-\xi|n|) \exp(\ii q n)$, with $q=2\pi/3$, $\xi=0.55$, $N=64$, $I_0/\tau=100$, $h=5\times10^{-5}$, $T_\textrm{end}=0.1$. Graphic velocity $V_{graph}\simeq 181$, $V_\textrm{teo}=\frac{2I_0}{\tau}\frac{\sinh(\xi)}{\xi}\sin(q)=182$, $H_{h,teo}=-2I_0\cosh(\xi)\cos(q)=0.138$,  $H_{h,num}(0)=0.104$.  {\bf (Right)} FFT of $c_n$ together with the theoretical phonon band $\omega=-\frac{2I_0}{\tau}\cos(q)$. $H_h$ losses 0.1\% of energy to the lattice after $T_\textrm{end}$.  Note: if $I_0/\tau$ diminishes the charge frequencies became smaller than the lattice ones and the charge becomes blocked leaving the energy to the lattice.
 }%end caption
\label{fig_localized}
\end{figure}

\section{Simulation tests}
\label{sec:simulationtests}
In this section, we test some physically interesting initial solutions and observe the result of the integration of the full system and some of its properties to check the model proposed. We limit ourselves to simulations for an extra hole, that is, $Q=+1$ in the previous sections.

The preferred numerical methods are  those that preserve the physical properties of the system at each integration step, in particular, charge probability conservation. %,  and are also symplectic, which guarantees that the numerical solution is close to the actual one.
They are described in \ref{sec:integrators}.
%INTEGRATOR

\subsection{Extended solutions}
Linear solutions of the linearized equations are extended ones $c_n=\frac{1}{\sqrt{N}}\exp(\ii [q n-\omega t])$. Substitution in \eqref{eq:unddotlinear} and \eqref{eq:cndotlinear} leads to $\ddot u_n=0$ and $\tau\omega=-2I_0\cos(q)$ and the charge Hamiltonian and frequency are $H_h=\tau\omega=E_0-2I_0\cos(q)$ and $\omega=E_0/\tau-2\frac{I_0}{\tau} \cos(q)$ as seen above.
  %"It can also obtained by substitution from the expected charge Hamiltonian in Eq.~(\ref{eq:H_h}).
The lattice Hamiltonian is zero because $u_n=0$ and $p_n=0$. Note that the unit energy in scaled variables is exactly $u_E=k_ce^2/a$, the electrostatic potential energy of a unit charge at the lattice unit distance. So an extra charge  would provide twice that amount if the charge is localized in a single site, but it is diminished for the extended solution. The velocity of the waves in $c_n$ should be the phase velocity as there is, in principle, a single plane wave, that is, $V_{teo}=\frac{\omega}{q}=-\frac{2}{q}\frac{I_0}{\tau}\cos(q)$.

\noindent The physical reason for the lattice to remain frozen is that the charge density is constant  because $|c_n|^2=1/N$ so each charge is subjected to opposite repulsive forces from each neighbor with the same modulus that cancel themselves, i.e.:
 $\ddot u=-Q|c_{n+1}^2|+Q|c_{n-1}^2|=-Q/N+Q/N=0$.
These solutions are somewhat irrelevant since nothing happens. However, they are very useful for coherence. %See Fig.\,\ref{fig_extended}

\subsection{Traveling localized trial functions}
We propose the trial function $c_n=A_0\exp(-\xi|n-V_bt|) \exp(\ii [q n-\omega t])$, which is $c_n=A_0\exp(-\xi[n-V_bt]) \exp(\ii [q n-\omega t])$ for $n-V_bt >0 $ and $c_n=A_0\exp(+\xi[n-V_b t])\exp(\ii[qn-\omega t])$ when  $n-V_b t<0$ alternatively changing $\xi$ for $-\xi$ when $n-V_b t<0$. It is easy to see that the time derivative is well defined in $n-V_b t=0$ and we will consider the coherence of this definition below. By substitution in Eq.~(\ref{eq:cndot}) and collecting together the real and imaginary coefficients we obtain:
\begin{eqnarray}
H_h=\tau\omega=-2I_0\cosh(\xi)\cos(q)\, ,\\
\tau\xi V_b=2I_0\sinh(\xi)\sin(q)\, .
\end{eqnarray}
We observe that changing $\xi$ for $-\xi$ does not change the above equations, and therefore they are valid for both tails of the trial function. Also, note that $V_b$ and $\sin(q)$ have the same sign. See Fig.\,\ref{fig_localized}.

The trial function $c_n$ is not a solution, and therefore it spreads. Simulation times should be the order of the theoretical period $T_\mathrm{teo}=2\pi/\omega$, with $\omega=H_h/\tau$ above.  However, they are a simple one of testing the equations, the simulation code and to get insight into the physics of the system. It is remarkable how well works for the tails of an actual solution.

\begin{figure}[b]
\includegraphics[width=0.49\textwidth]{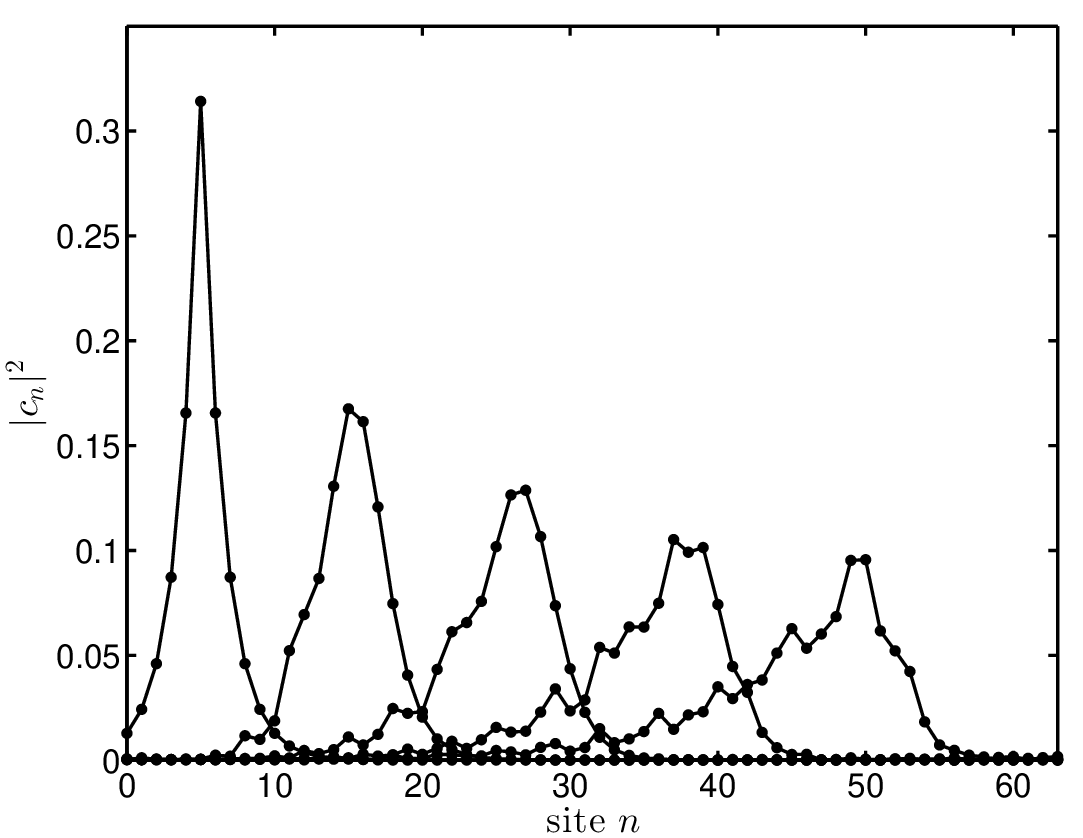}
\includegraphics[width=0.49\textwidth]{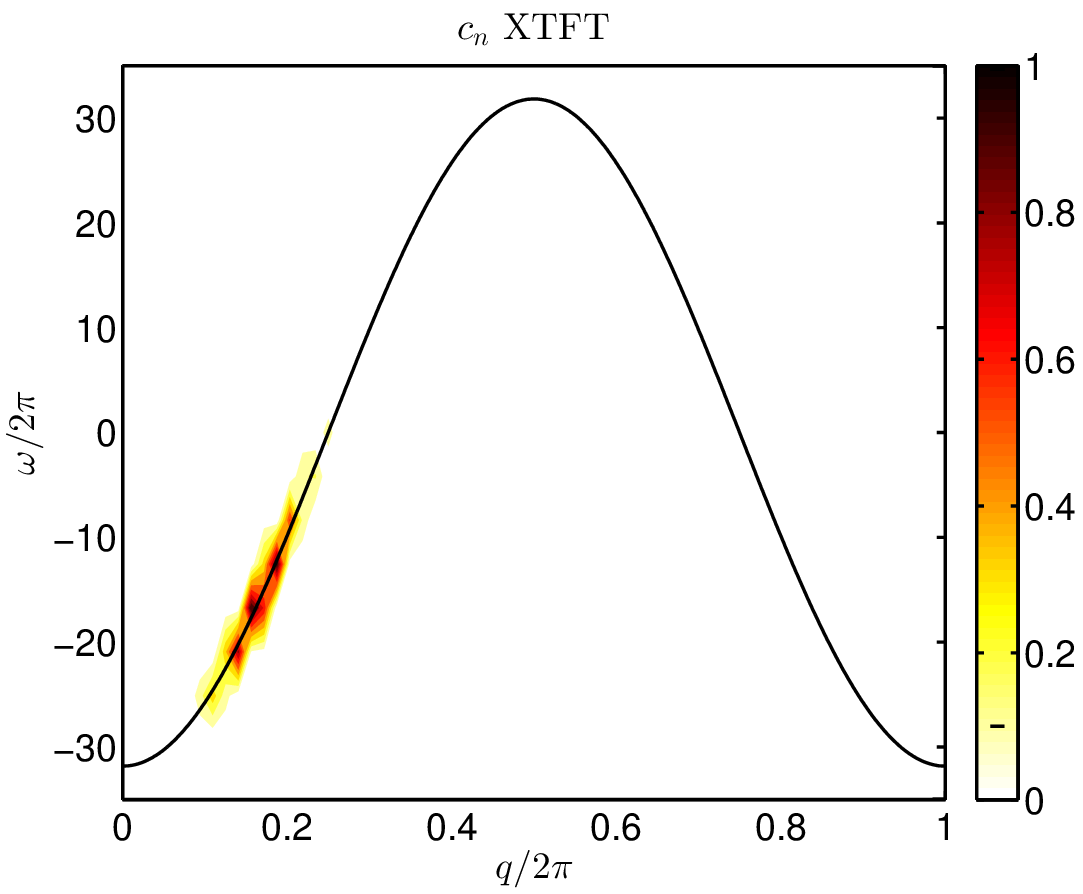}
\caption{{\bf (Left)} Localized wave obtained with initial conditions $u=0$, $p=0$, $c_n=A\exp(-\xi|n|) \exp(\ii q n)$, with $q=\pi/3$, $N=64$, $I_0/\tau=100$,
$\xi=0.32$.  Integration parameters: $h=6\times10^{-5}$,
  %$\xi=\mathrm{acosh}(\alpha J0 \cos(q))=0.32$, so as the interaction with the lattice disappears at 0th order. Integration parameters: $h=6\times10^{-5}$,
$T_\textrm{end}=0.23$ and 4000 steps. Results: graphic velocity $V_{graph}\simeq 186$, $V_\textrm{teo}=\frac{2I_0}{\tau}\frac{\sinh(\xi)}{\xi}\sin(q)=176$, $H_{h,teo}=-2I_0\cosh(\xi)\cos(q)=-0.126$,  $H_{h,num}(0)=-0.1131$.  {\bf (Right)} FFT of $c_n$ together with the theoretical phonon band $\omega=-\frac{2I_0}{\tau}\cos(q)$. Note: $H_h$ loses 0.006\% of energy to the lattice after $T_\textrm{end}=4T_\mathrm{teo}$.
The trial solution spreads in despite the small interaction with the lattice due to the hopping probability.
 }%end caption
\label{fig_localnolatt1}
\end{figure}

\begin{figure}
\includegraphics[width=0.49\textwidth]{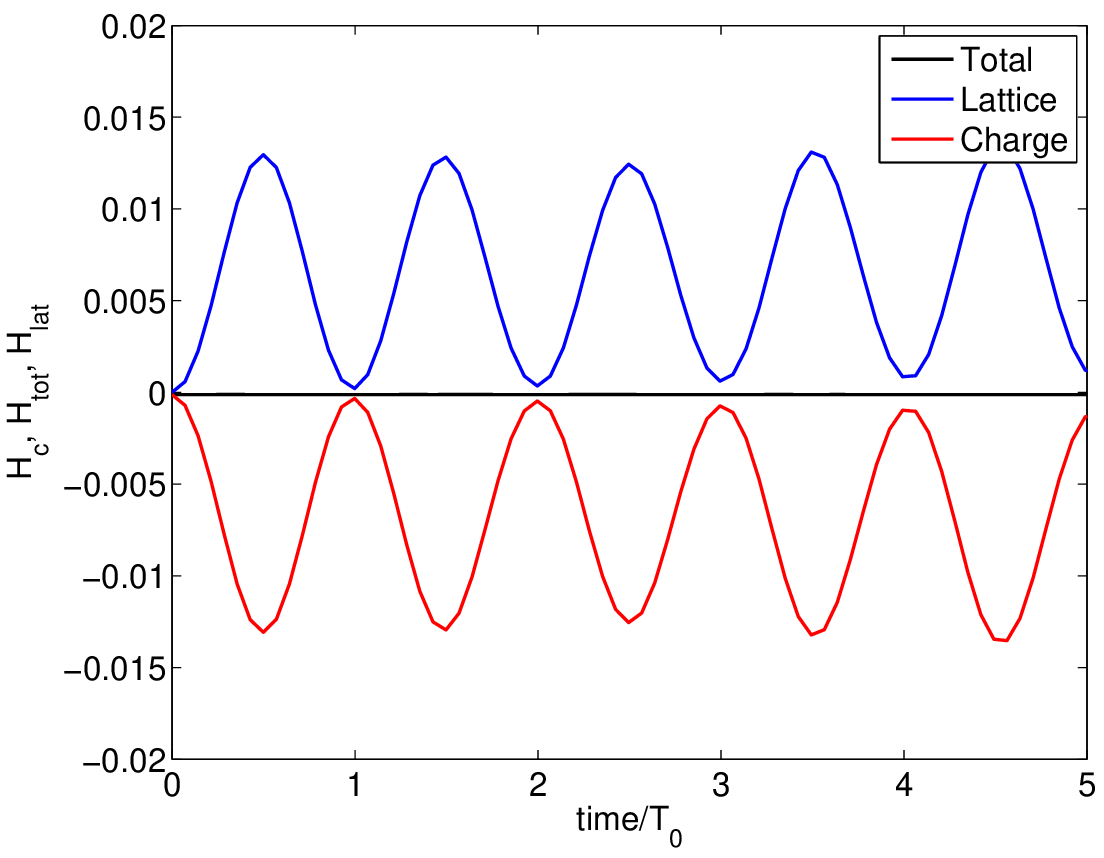}
\includegraphics[width=0.49\textwidth]{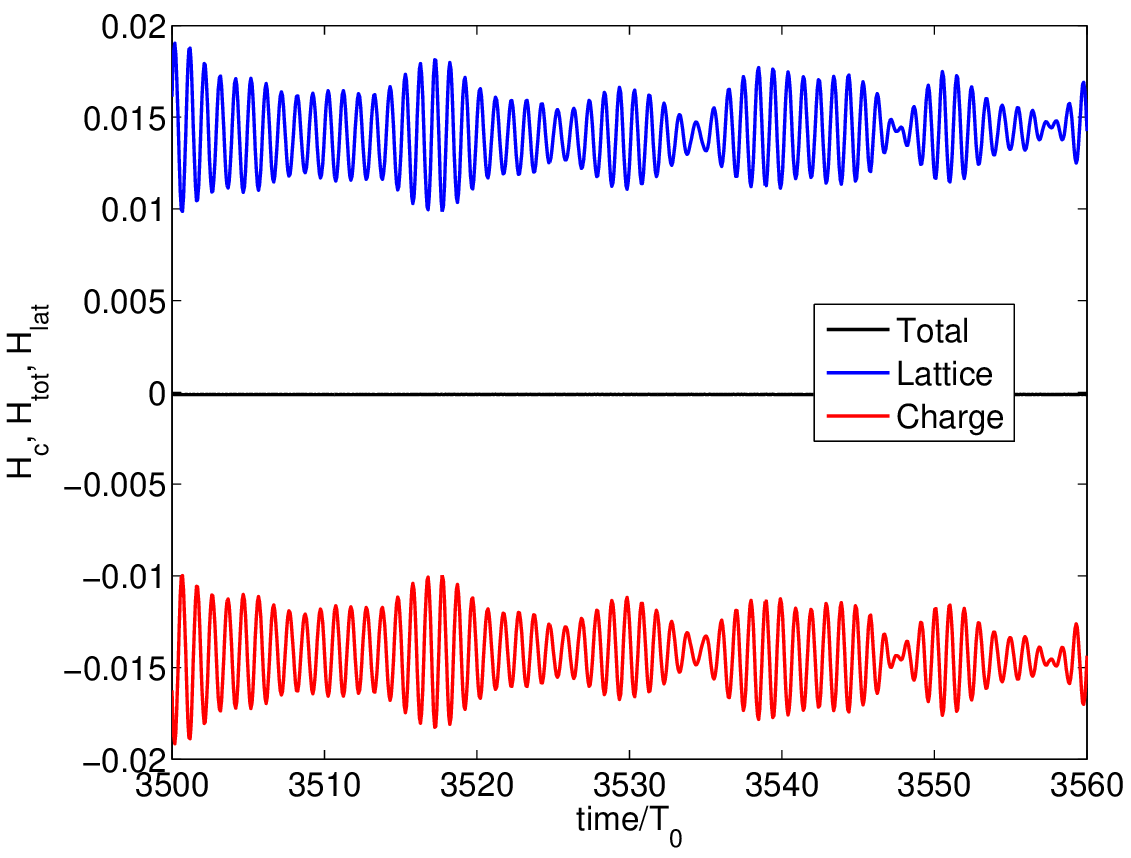}
\caption{Energy evolution obtained with initial conditions $u=0$, $p=0$, $c_n=A\exp(-\xi|n|) \exp(\ii q n)$, with $q=\pi/3$, $N=64$, $I_0/\tau=0.1127$, $\alpha=12.45$ and integration step: $h=0.01$. $T_0=1.4$ is the period of decoupled lattice small oscillations. The lattice gets very quickly energy from the charge. {\bf (Left)} First 5 periods with $T\simeq T_0$. {\bf (Right)} Last 60 periods with the same main period and mean charge/lattice energy, corresponding to an energy charge of 0.015 or 40\,meV.  The process is accompanied by a small increase in localization and a moderate rupture of the monotony of the decreasing pattern. }%end caption
\label{fig_localnolatt2}
\end{figure}

\subsection{Stationary localized trial functions}
There are two stationary trial functions $c_n$: if $V_b=0$, $\sin(q)=0$ and $q=0$ and $q=\pm\pi$ (same physical wavevector). For $q=0$, $H_h=-2I_0\cosh(\xi)$ and for $q=\pm\pi$, $H_h=2I_0\cosh(\xi)$.

\begin{figure}
\includegraphics[width=0.49\textwidth]{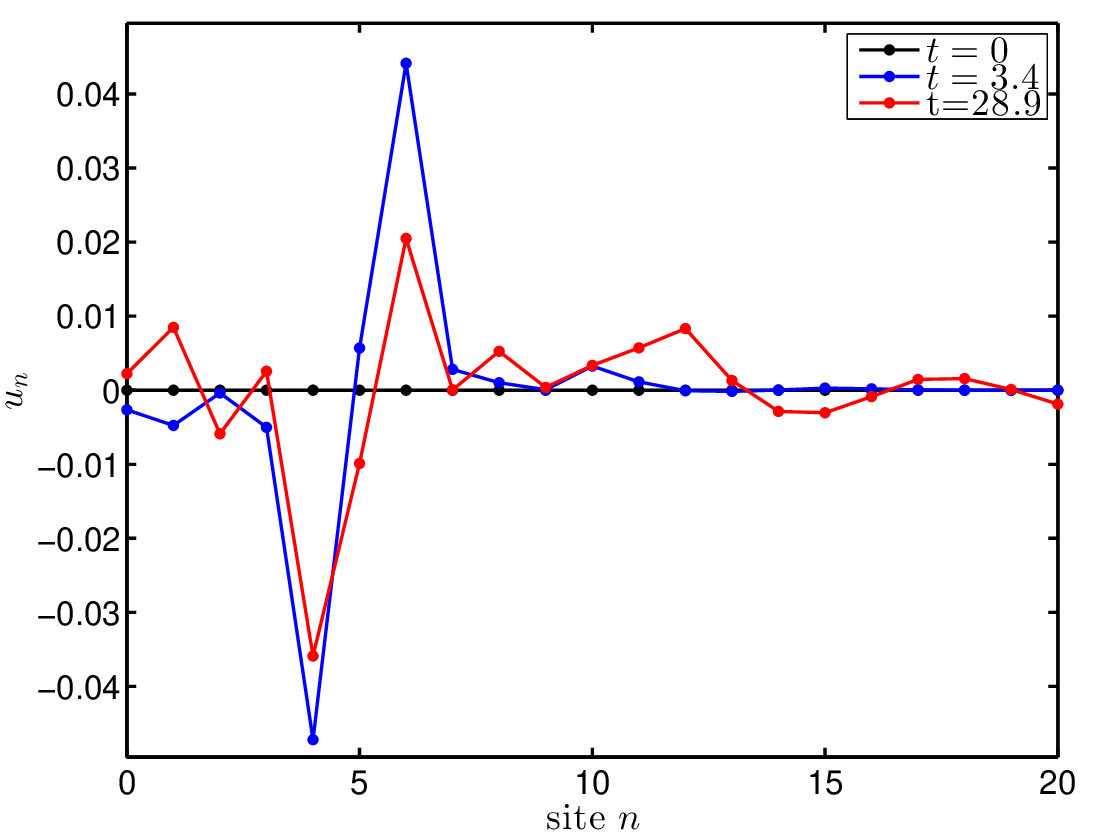}
\includegraphics[width=0.49\textwidth]{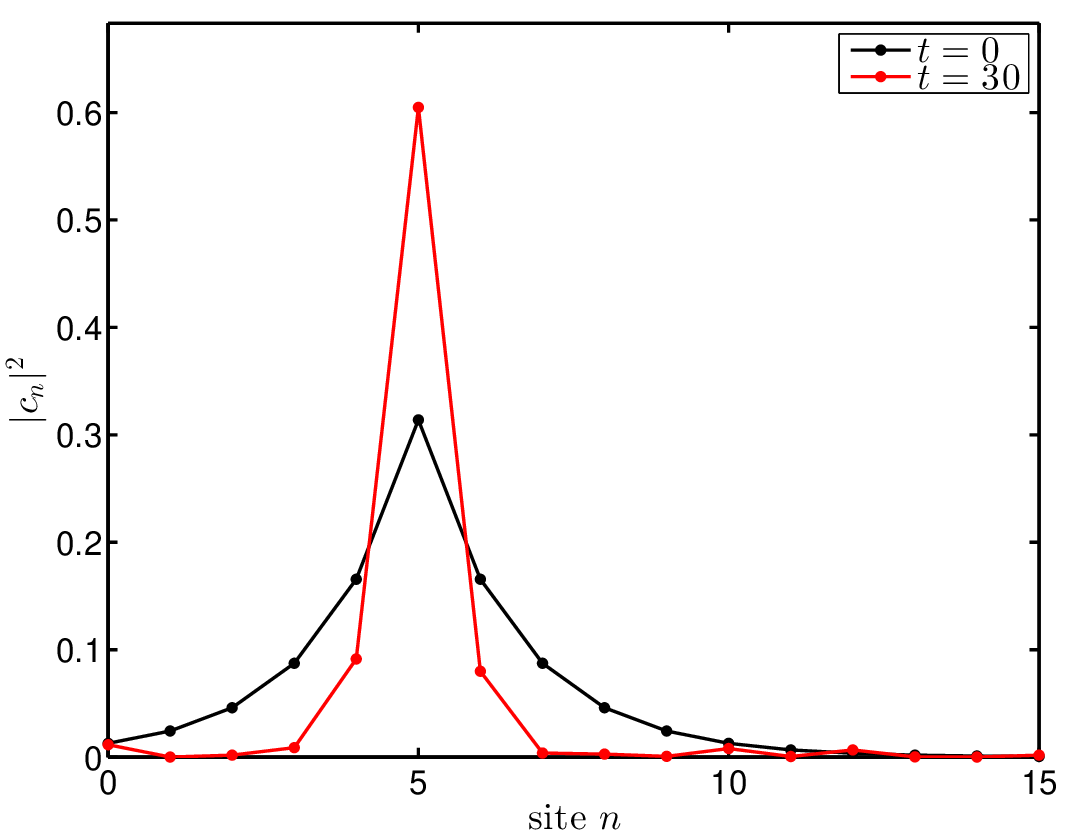}
\caption{{\bf (Left)} Approximated oscillating self-localized mode obtained with initial conditions $u=0$, $p=0$, $c_n=A\exp(-\xi|n|) \exp(\ii q n)$, with $q=0$, $N=64$, $I_0/\tau=1$, $\xi=0.32$. Integration parameters: $h=1.4\times10^{-3}$, $T_\textrm{end}\simeq 30$ and 213090 steps. Results:  $H_{h,teo}=-2I_0\cosh(\xi)=-0.0252$,  $H_{h,num}(0)=-0.00226$, then it oscillates between -0.02 and -0.05 with corresponding oscillations of the lattice Hamiltonian.  {\bf (Right)} Self-localization of the charge density.
 }%end caption
\label{fig_staq0}
\end{figure}

For large values of $I_0/\tau$ as 100 or 10, the charge probability spreads rapidly, for values as  $I_0/\tau=1$ the lattice couples with the lattice and for $T_h=2.97$ larger than $T_0=1.4$, meaning that the lattice evolves faster than the charge. Interestingly, there is the phenomenon of self-localization, a localized vibration of $u_n$ develops at $n_0$, the particle with more charge at $t=0$, affecting the two neighbors and at the same time the charge probability becomes more concentrated in that same particle than at the beginning, as can be seen in Fig.\,\ref{fig_staq0}. %\comment{This happens to any $q$}

\label{sec:someimulations}
\subsection{Stationary charge}
\label{subsec:stationarycharge}
%We have performed some simulations to study the coherence of the system, using $I_0=0.01\tau$ to $I_0=10\tau$

The simplest initial conditions are provided by the lattice at equilibrium $u=0$, $p=0$, and the location of a charge at site $n$, i.e., $|c_{m}|^2=\delta(n,m)$,  with $a_n=1,b_n=0$. Any other combination of $a_n$ and $b_n$ that keeps the probability is equivalent.

\begin{figure}[t]
\begin{center}
\includegraphics[width=\imagesize]{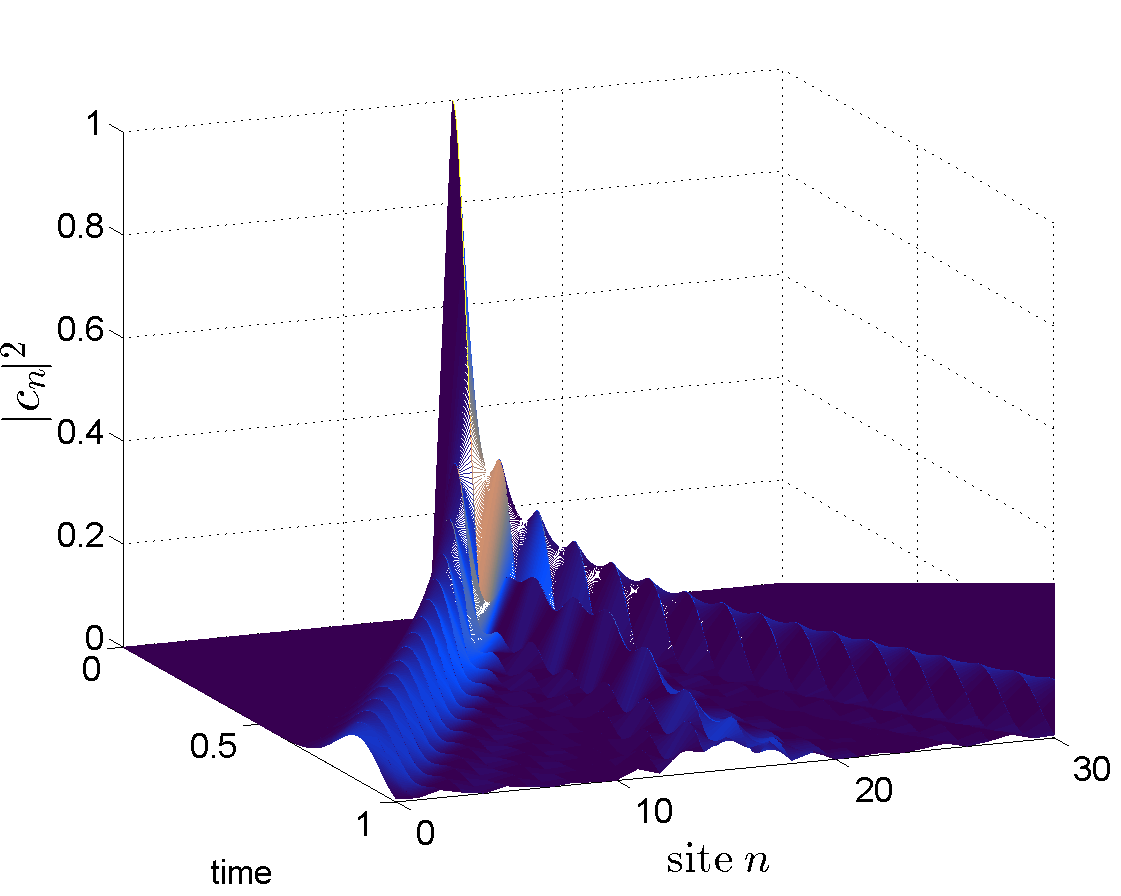}
\end{center}
\caption{Charge spread in a lattice initially at rest, $c_{16}=1$.  Parameters: $h=10^{-3}$, $N=1000$, $I_0/\tau=10$. The probability is divided into three. Any of the traveling probabilities corresponds
to traveling charge without much perturbation of the lattice.
 }%end caption
\label{fig_chargespread}
\end{figure}

We observe that the charge probability does not spread until quite high values of $I_0/\tau$, actually to obtain a fast spread we need $I_0/\tau=10$. This is coherent with the properties of muscovite actually being an insulator. Figure\,\ref{fig_chargespread} shows this spread.
\begin{figure}[h]
\begin{center}
\includegraphics[width=\imagesize]{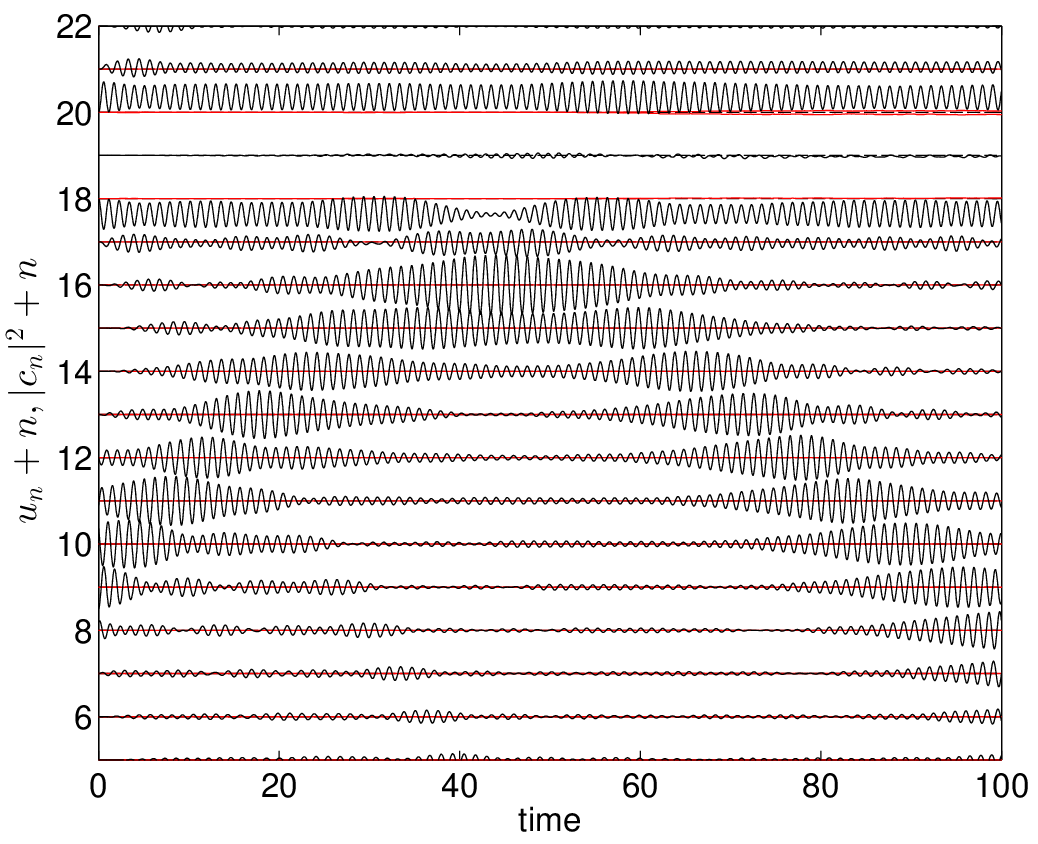}
\end{center}
\caption{Approximate breather generated with $[u_9, \dots, u_{13}]$ = [0.0183, -0.0501, 0.0370, $-0.0106$, 0.0011], $[p_9, \dots, p_{13}]$ = [0.0692, 0.0056, -0.1636, 0.1448, -0.0848], $c_{20}=1$. The charge $c_n$ is in red, and the lattice coordinates are amplified 10 times.   Parameters: $h=0.01$, $N=10000$, $I_0/\tau=0.1$.
 }%end caption
\label{fig_breatherrebounds}
\end{figure}

\subsection{Breather rebounding in a charge}
With a simple pattern, it is possible to produce non-exact breathers. If we locate an extra charge in their vicinity, the breather rebounds, while the charge keeps its localized position. Initially, a symmetrical oscillation of the neighboring particles to the charge develops. See Fig.\,\ref{fig_breatherrebounds}.

\subsection{Kink with an extra charge}
\begin{figure}[h]
\begin{center}
\includegraphics[width=0.49\textwidth]{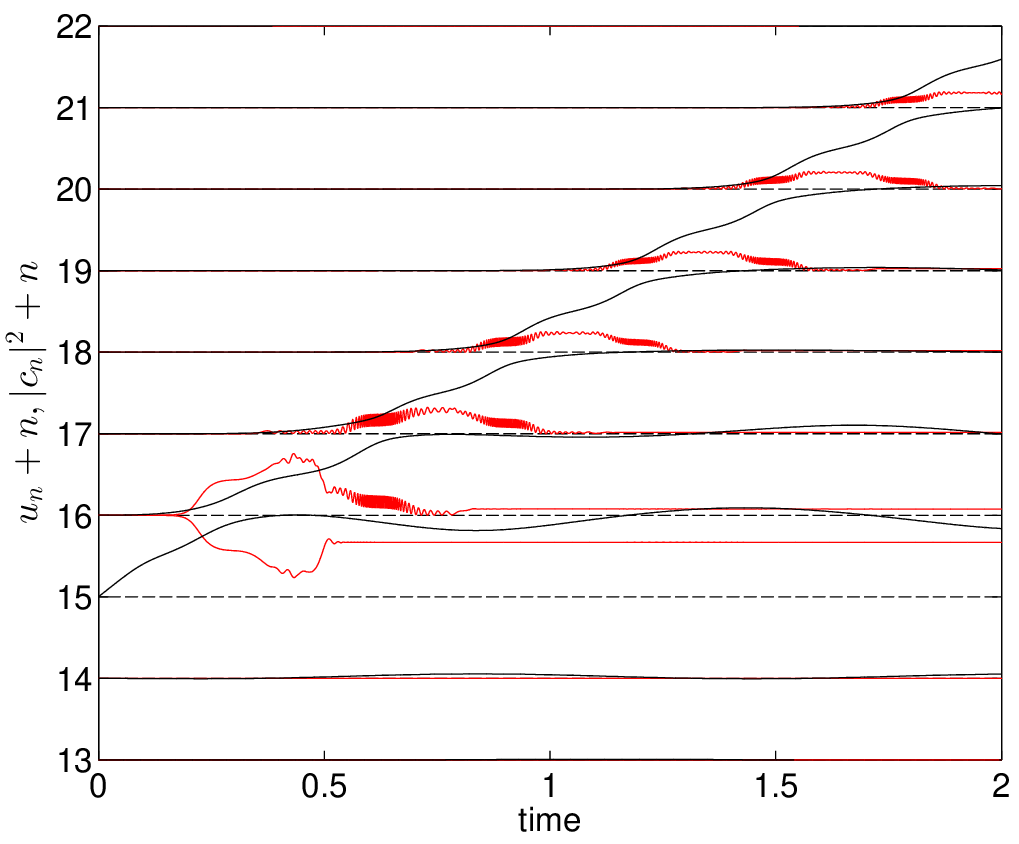}
\includegraphics[width=0.49\textwidth]{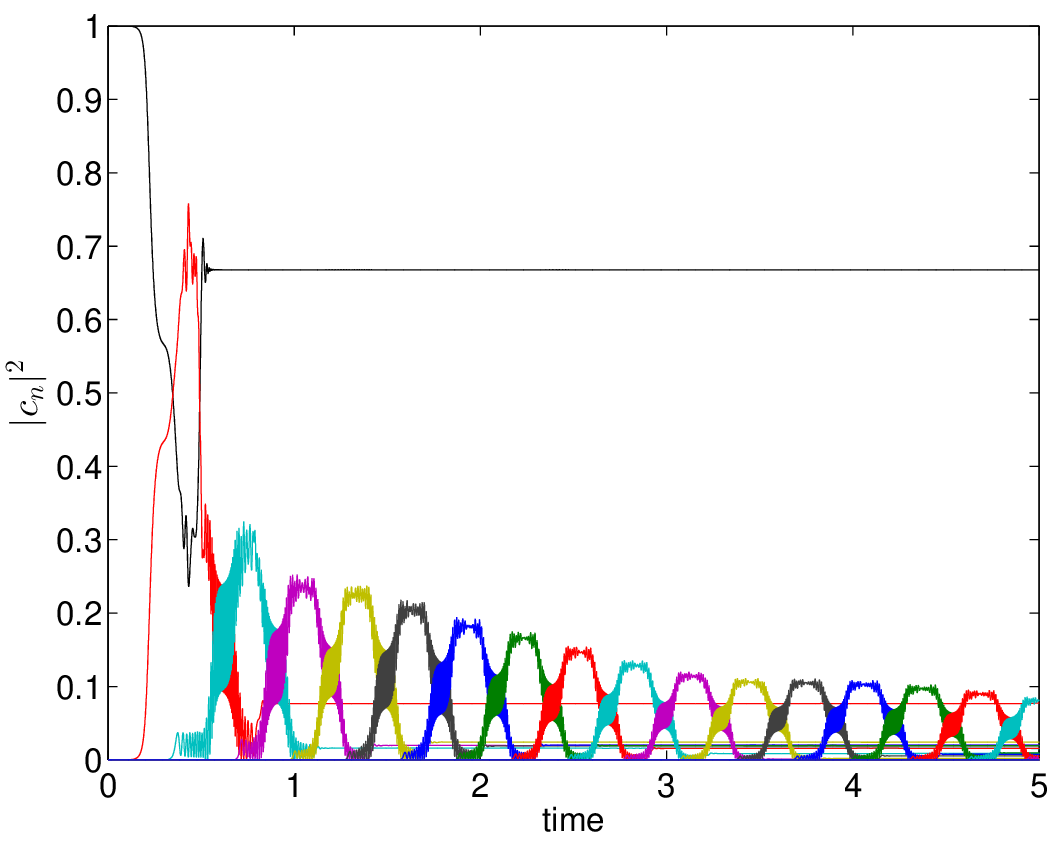}
\end{center}
\caption{{\bf (Left)} A kink produced with initial momentum   %$p(n)=6,   Ek=49.9(\,eV) and charge $A(n)=1$
$p_{15}=4.4$ ($E_K=0.5p^2$=26.8\,eV) and a localized charge with $c_{15}=1$.  A kink is produced and the charge probability density $|c_n|^2$ (in red)
is partially left at the initial particle and partially travels with the kink. Parameters: $h=10^{-4}$, $I_0/\tau=0.01$, $T_\mathrm{end}=5$ and $N=64$ particles. Dashed lines are reference lines both for the charge and the lattice variables. Results: $\simeq 27$\% probability is carried initially with the kink, diminishing to $\simeq 9$\% at time $t=5$. A stable probability of 67\% is left at the initial site. {\bf (Right)}  A small amount of charge is left stable with each particle after the passage of the kink. For $I_0/\tau=0.1$, the same kick does not produce a kink, for a larger kick of $p_{16}=6$, it does. A smaller probability of 12\% travels with the charge but with no observable dispersion.  A larger value of $I_0/\tau$ needs studying as there are bursts of increase of the lattice Hamiltonian and (negative) charge Hamiltonian, although the total Hamiltonian is conserved.
 }%end caption
\label{fig_badkink}
\end{figure}

Note that for kinks, the particles became very close, and the energy can change very rapidly, therefore, a smaller step $h$ might be necessary. Kinks are produced without charge at energies of 26.2 eV\,\cite{archilla-kosevich-pre2015,archilla-zolotaryuk2018}, with $u_E\simeq 2.77$\,eV, the velocity to be provided to a single particle should be about $V_b=\sqrt{2\times 26.2/2.77}\simeq 4.35$ in scaled units. Locating the charge with $c_{15}=1$ and $p_{15}=4.4$ with $I_0/\tau=0.01$, we can test the system. With step $h=0.001$, the energy is not conserved while the charge is always conserved due to our numerical method. So we use $h=10^{-4}$ and obtain a kink for the lattice variables. The charge probability is divided. One part 67\% is located at the initial particle and a smaller one travels with the kink, as can be seen in Fig.\,\ref{fig_badkink}. There is always a small probability left at each particle and lost to the kink.

The process depends heavily on $I_0/\tau$. For $I_0/\tau=0.1$, both the charge and the lattice vibration remain trapped, but increasing the initial momentum to $p_{15}=6$, the kink reappears traveling with 12\% probability with the charge. In this case, the charge is not dispersed, which is more favorable than $I_0/\tau=0.01$. %See {\em measurements} below.

Note that in an ionic crystal, the movement of an ion implies the movement of electric charge by itself. In this case, the charge transported is larger $+2e$. This might be coherent with the thick lines of primary quodons.
%{\em Not with probability one}, $|c_n|^2$ is a probability, not a charge density.

%\noindent {\bf Measurements}
%We can perform a {\em measurement}, that is restart the simulation from a given time,so as the charge is located at the kink. Then re-start the simulation.
%It is not clear if the state of the system is altered. But, if it is so, that is coherent with quantum mechanics, where a measurements produces a change of the system known as the collapse of the wave function $\{c_n\}$ because the charge has been located at some specific points. We will explore this phenomenon in future publications.

\subsection{Chaotic breather with an extra charge}
\label{sec:chaobreatherwithcharge}
\begin{figure}
\begin{center}
\includegraphics[width=\doublefig]{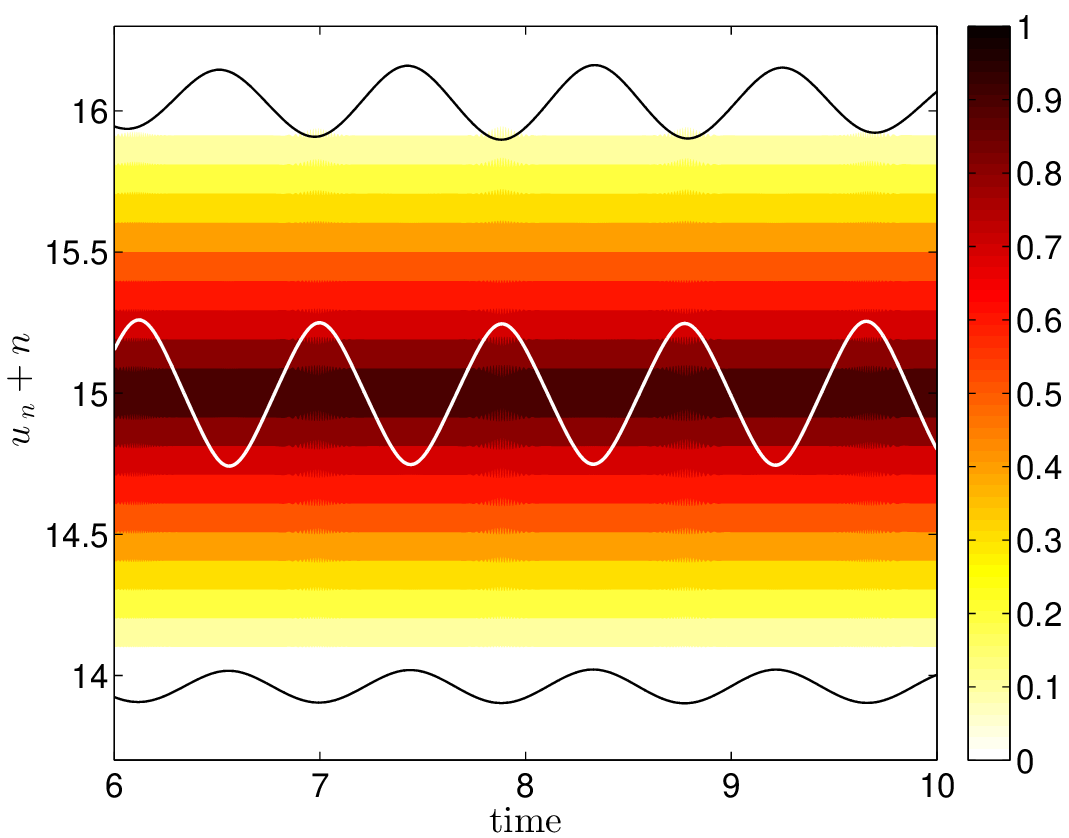}\mbox{}
\includegraphics[width=\doublefig]{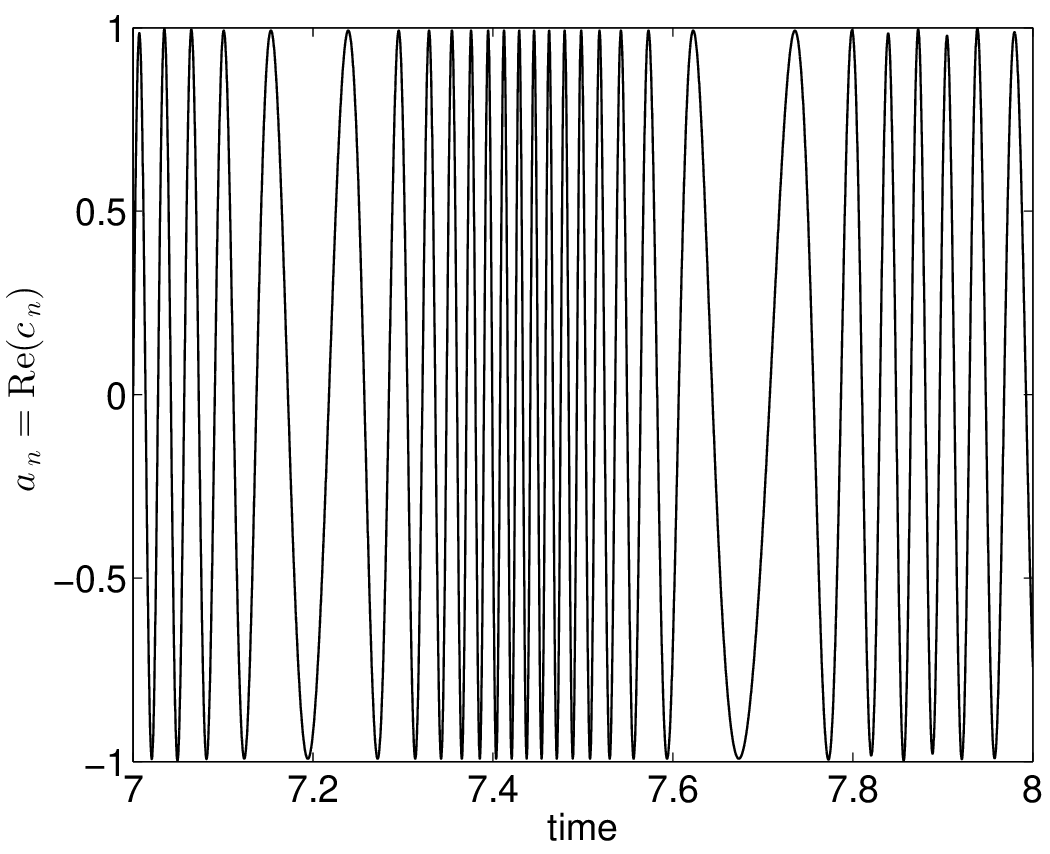}
\end{center}
\caption{Quasi-periodic chaotic breather with an extra charge. {\bf (Left)} The coordinates of particle 15 and the two nearest neighbors are shown with apparent but not exact periodicity. The charge probability represented as color remains localized in the initial particle after an initial small spread at the nearest neighbors. {\bf (Right)} The real part of the charge amplitude shows the chaotic behaviour with $a_n\simeq 1$ repeating but non periodically.
Parameters $I_0/\tau=0.1117$, $\alpha=12.45$, integration step $h=10^{-5}$.
  }%end caption
\label{fig_quasibreathercoordinates}
\end{figure}

\begin{figure}
\begin{center}
\includegraphics[width=\doublefig]{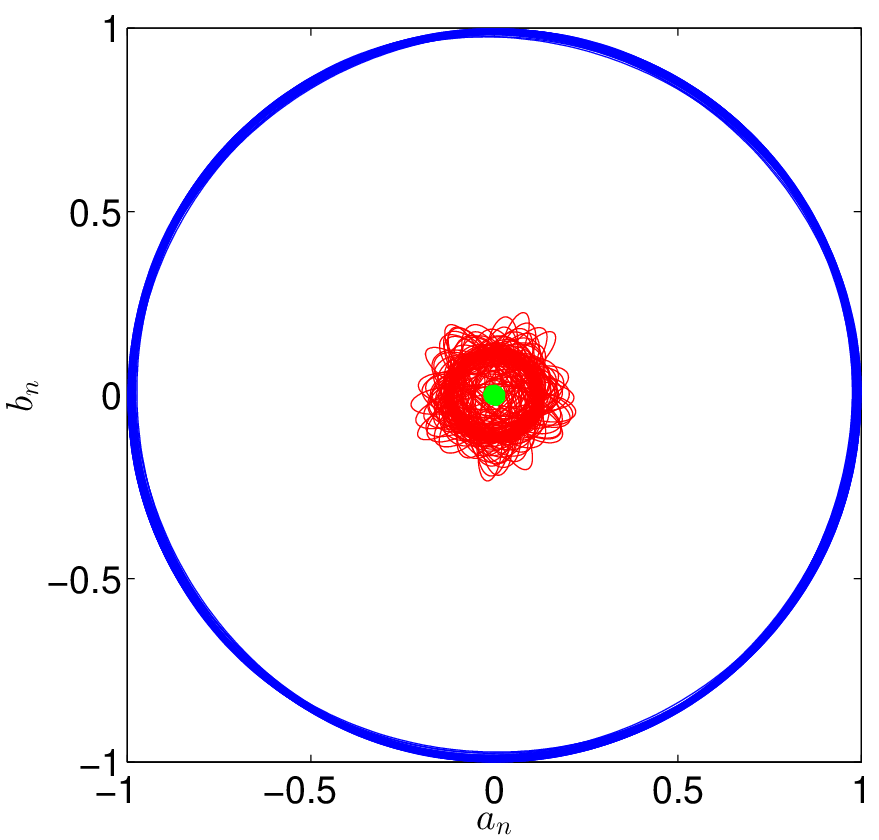}
\includegraphics[width=\doublefig]{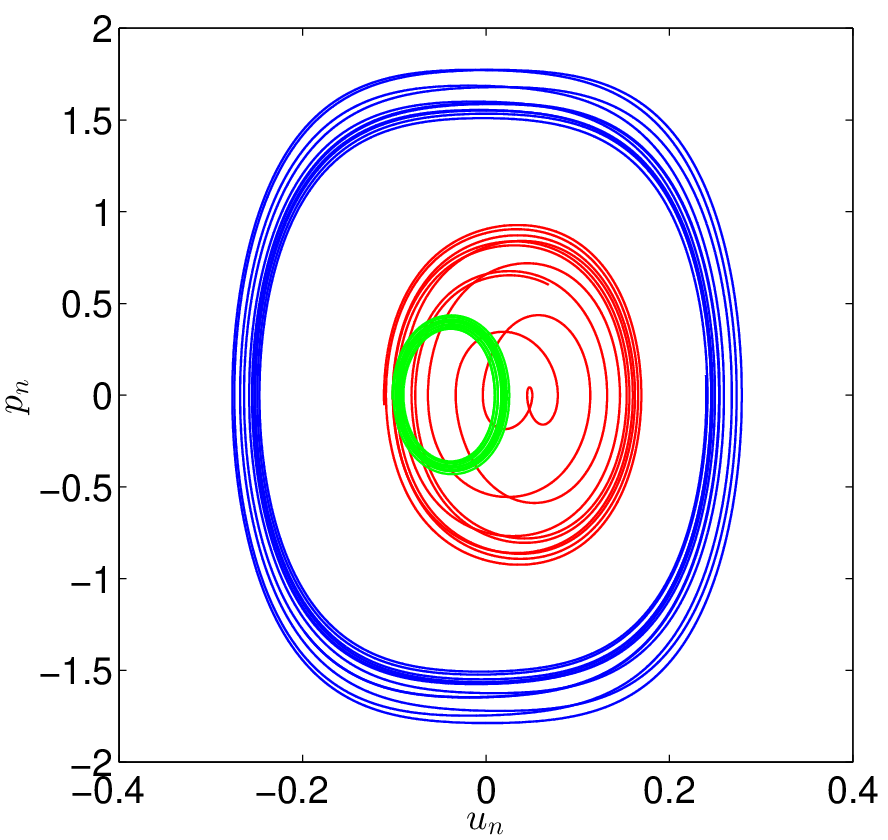}
\end{center}
\caption{Phase space of the phase spaces of the chaotic breather with an extra charge as described in Section\,\ref{sec:chaobreatherwithcharge}.   In blue, the core particles and, in red and green, the two nearest neighbors. {\bf (Left)} Charge amplitude real and imaginary part $a_n$ and $b_n$.
{\bf (Right)}  Lattice coordinate and momentum $u_n$ and $p_n$. See text.
Parameters $I_0/\tau=0.1117$, $\alpha=12.45$, integration step $h=10^{-5}$.
  }%end caption
\label{fig_quasibreatherphases}
\end{figure}

We have found chaotic breathers\,\cite{ikeda-doi2007} with an extra charge that are quasi-periodic in the lattice variables and also for the charge amplitude or probability. This is an interesting possibility as it is a mechanism for trapping energy and charge during certain times. The pattern is close to the Page mode, that is, with a site with maximum amplitude with nearest neighbors with smaller amplitude, and opposite phase. This is a breather with high energy 4.5\,eV, 5\,eV corresponding to the lattice and -0.5\,eV to the charge. It is presented in Fig.\,\ref{fig_quasibreathercoordinates}-Left with the  parameters $\alpha=12.5$ and $I_0/\tau=0.1117$. The particle with initial probability one loses around 0.015, and then there is a small interchange to neighboring particles that is recovered but in a nonperiodic way, as it can be seen in Fig.\,\ref{fig_quasibreathercoordinates}-Right. The phase space of the variables $a_n$ and $b_n$ of the charge amplitude and the lattice displacements and momenta are also represented in Fig.\,\ref{fig_quasibreatherphases}. Only the core particle variables and the nearest neighbors are represented for clarity. This entity is called a chaotic breather or chaobreather for short. From the physical point of view it represents a different form of energy and charge localization, as observed in the hyperconductivity experiments described in Sect.\,\ref{sec:introduction}, where alpha particles initially bring about a current peak showing that a reservoir of charge has been mobilized.

\newpage
\section{Conclusions}
\label{sec:conclusions}
In this work, we have presented a model for charge transport along K$^+$ chains in silicates mediated by nonlinear excitations. The motivation for this work is the experimental observation of hyperconductivity, the phenomenon of charge transport in the absence of an electric field when a side of the crystal is bombarded with alpha particles. The model relies heavily upon the work of previous publications, but for the first time, an electric charge is considered through a quantum Hamiltonian. The vibronic interaction between ions and an extra electron or hole, described by the transfer integral, is strongly nonlinear, increasing as neighboring ions become closer enhancing the probability of charge transmission. This allows for the existence of localized charge states that break the discrete translational invariance of the lattice. These states, when mobile, are the proposed charge carriers in hyperconductivity experiments.
Most of the parameters are obtained through physical deduction, although some are yet not well known, particularly the transfer integral. Extensive work on that subject is being done, and it will be published elsewhere. In this work, we analyze the coherence of the model, its behavior, and spectra, for different initial conditions based on different ans\"atze, obtained from the tail analysis of extended and exponentially localized profiles, isolated charges, and other means. We have found interesting phenomena, such as the self-localization of some stationary solutions and the trap of a charge by a chaotic breather. The obtention of exact traveling solutions is the object of the present research, as well as the estimation of the missing parameters. Although in this work we have developed the model both for holes and electrons, we have limited the simulations to holes as they are the best candidate for the phenomenon of hyperconductivity because 99\% of the $^{40}$K decay leaves a positive charge behind. New development may require a modification or refinement of our model, but, at present, it seems physically sound. At present, the propagation of charge has not been achieved due to the large number of frequencies that appear due to the different amplitudes of the particle oscillations. We expect to solve this problem both numerically and conceptually for the physical system.  That work will be reported in future publications.

\section*{Acknowledgments}
%JFRA  thanks projects MICINN PID2019-109175GB-C22 and VII PPIT-US 2023.
%He also acknowledges the Universities of Osaka and Latvia for hospitality.
%JB acknowledges support from PostDocLatvia grant No.1.1.1.2/VIAA/4/20/617.
%YD  acknowledges the support from grant JSPS Kakenhi (C) No. 19K03654.
%MK acknowledges support from grants  JSPS Kakenhi (C) No. 21K03935.
JFRA  thanks projects MICINN PID2019-109175GB-C22, PID2022-138321NB-C22, and VII PPIT-US 2024.
He also acknowledges the Universities of Osaka and Latvia for hospitality.
JB acknowledges financial support by the Faculty of Physics, Mathematics and
Optometry of the University of Latvia.
YD acknowledges JSPS Kakenhi (C) No. 19K03654.
MK acknowledges support from JSPS Kakenhi (C) No. 24K07393 and 21K03935.

\section*{References}
\bibliography{osaka2023JT} %{russell2021}
\appendix
\section{Real canonical Hamiltonian  equations}
\label{sec:hamiltonianODE}
The equations in Section\,\ref{subsec:finalequations}  can be converted into canonical Hamiltonian equations for the non-separable Hamiltonian $H=H_\textrm{lat}+H_h$, with  $H_\textrm{lat}$ and $H_h$, given by Eqs.~(\ref{eq:Hlat})--(\ref{eq:H_h}). Let us denote $a_n$ and $b_n$ as the real and imaginary part of $c_n$, and let us define the scaled variables $A_n=\sqrt{2\tau} a_n$, $B_n=\sqrt{2\tau} b_n$, then $B_n$ is the conjugate momentum of $A_n$. The coordinate variables become $z=[u_1,\dots,u_N,A_1,\dots,A_N]$ and the momenta variables $\Pi=[p_1,\dots,p_N,B_1,\dots,B_N]$.

With this notation:
\begin{eqnarray}
\dot u_n=\frac{\partial H}{\partial p_n}; \quad \dot A_n=\frac{\partial H}{\partial B_n};\quad
\dot p_n=-\frac{\partial H}{\partial u_n}; \quad \dot B_n=-\frac{\partial H}{\partial A_n}\,.
\end{eqnarray}
Or
\begin{eqnarray}
\dot z_n=\frac{\partial H}{\partial \Pi_n}; \quad \dot \Pi_n=-\frac{\partial H}{\partial z_n}.
\end{eqnarray}

This system is real and in the form of canonical Hamiltonian equations, and it is convenient for numerical integration.

With this scaling, the new evolution equations for the displacements are changed, while the equations for the charge show no change:
\begin{eqnarray}
&\mbox{}&\dot p_n=-U'(u_n)-Q U'_h(u_n)\frac{(A_n^2+B_n^2)}{2\tau}\nonumber\\
&\mbox{}& -\frac{1}{(1+u_{n+1}-u_n)^2}\left[ 1+C\exp(-\beta(u_{n+1}-u_n))+Q\frac{(A_n^2+B_n^2+A_{n+1}^2+B_{n+1}^2 )}{2\tau}  \right]\nonumber\\
&\mbox{}&+\frac{1}{(1+u_{n}-u_{n-1})^2}\left[ 1+C\exp(-\beta(u_{n}-u_{n-1}))+Q\frac{(A_n^2+B_n^2+A_{n-1}^2+B_{n-1}^2)}{2\tau}   \right]\nonumber\\
&\mbox{}&+\frac{C\beta\exp(-\beta(u_n-u_{n-1}))}{1+u_n-u_{n-1}}-
\frac{C\beta\exp(-\beta(u_{n+1}-u_n))}{1+u_{n+1}-u_{n}}\nonumber \\
&\mbox{}&+2\alpha I_0\exp(-\alpha(u_{n+1}-u_n))\frac{(A_{n+1}A_n+B_{n+1}B_n)}{2\tau}\nonumber\\
&\mbox{}&-2\alpha I_0\exp(-\alpha(u_{n}-u_{n-1}))\frac{(A_{n}A_{n-1}+B_{n}B_{n-1})}{2\tau}\, ,
\label{eq:unddotReAnBn}
\end{eqnarray}
and:
\begin{eqnarray}
\dot A_n=\frac{1}{\tau}\left[Q U_h(u_n)+\frac{Q}{1+u_n-u_{n-1}}+\frac{Q}{1+u_{n+1}-u_n}-2Q+E_0\right]B_n\\
 -\frac{I_0}{\tau}\exp(-\alpha(u_n-u_{n-1}))B_{n-1}- \frac{I_0}{\tau}\exp(-\alpha (u_{n+1}-u_{n})) B_{n+1}\,;\nonumber\\
\dot B_n=-\frac{1}{\tau}\left[QU_h(u_n)+\frac{Q}{1+u_n-u_{n-1}}+\frac{Q}{1+u_{n+1}-u_n}-2Q+E_0\right]A_n\nonumber\\
+\frac{I_0}{\tau}\exp(-\alpha[u_n-u_{n-1}])A_{n-1}+\frac{I_0}{\tau}\exp(-\alpha [u_{n+1}-u_{n}]) A_{n+1}\,.
\label{eq:AnBndot}
\end{eqnarray}

The lattice Hamiltonian remains unchanged, and the charge Hamiltonian changes in the obvious way. We reproduce it here for completeness:
\begin{eqnarray}
H_{lat}&={\displaystyle \sum_n \frac{1}{2}p_n^2+U(u_n)+V(u_{n}-u_{n-1})}\,,  \quad \textrm{with}\\
 V(u_{n}-u_{n-1})&=\fracc{1}{1+u_n-u_{n-1}}+\fracc{C}{1+u_n-u_{n-1}}\exp(-\fracc{u_n-u_{n-1}}{\rho})\,.
 \label{eq:Hlat2}
 \end{eqnarray}
 For energy calculations, it will be convenient to subtract $(1+C)=V(0)$ from $V$, i.e., so as the energy is zero at equilibrium, but we do not include it here for clarity and because the dynamical equations do not depend on a constant term.

 The expected value of the charge Hamiltonian in a generic state $\ket{\phi}=\sum_n c_n |n\rangle $ is given by:
 \begin{eqnarray}
 H_Q = \sum_n E_n\frac{(A_n^2+B_n^2)}{2\tau}-2 J_{n,n+1}{\tau}\frac{A_n A_{n+1}+B_n B_{n+1}}{2\tau}\, ,
\label{eq:H_QAnBncompact}
\end{eqnarray}
or explicitly:
\begin{eqnarray}
H_Q&=& \sum_n  \left(QU_h(u_n)+ \frac{Q}{1+u_n-u_{n-1}}+\frac{Q}{1+u_{n+1}-u_n}-2Q+E_0 \right)\frac{(A_n^2+B_n^2)}{2\tau}\nonumber\\
&-& 2 I_0 \exp(-\alpha(u_{n+1}-u_{n})) \frac{(A_n A_{n+1}+B_n B_{n+1})}{2\tau}\, .
\label{eq:H_QAnBn}
\end{eqnarray}

With this change of variables, we have obtained canonical Hamiltonian equations, but also, the system is a bit more amenable as the scaling has moved from the time to the probability amplitudes $(A, B)$.  As seen in Sect.\,\ref{sec:linear},  the parameter $\Omega_Q=2 I_0/\tau$ is the maximum linear frequency of the charge of the order of magnitude of the lattice minimal frequency.  The eigenfrequency $E_n/\tau$ should be relatively small for breathers because  $E_n$ is zero at equilibrium and, therefore, small for the relatively small displacements corresponding to breathers.

\section{Numerical integration}
\label{sec:integrators}

In this section, we use it as the main reference Ref.\,\cite{hairerbook2006}. Our system of ordinary differential equations (ODE) is canonical Hamiltonian, and  most important, it  conserves the charge probability, which is a property of the Schr\"odinger equation. Therefore, it is suitable to construct a numerical integration scheme that holds the same important physical properties.
 %However, generically speaking, it is not possible for a numerical integrator to conserve both the Hamiltonian and to be
 % symplectic\,\cite{zhongmarsden1988,sanzsernabook1994}.
%It is preferably to conserve symplecticity which have a good approximate conservation of the Hamiltonian and produce very accurate solutions with respect to the ODE. Note that the fact that the Hamiltonian is conserved does not guarantee that the solution is accurate.
%The exception to this incompatibility are Hamiltonians or other properties that are quadratic and symmetric. In this case, there are integrator method %that conserve those properties. We will use this property in our integration method.

Another interesting properties of integrators is whether they are explicit or implicit. Explicit methods are  much more efficient, but they are possible only when the Hamiltonian is separable, i.e., it can be written as a sum of parts with only generalized coordinates, but our Hamiltonian is not separable: $H\neq H(\Pi)+H(z)$.

With these limitations, let us write the equations of motion  in condensed  form to clarify their structure:
\begin{eqnarray}
\dot u&=&\frac{\partial H}{\partial p}=p\, \\
\dot A&=&\frac{\partial H}{\partial B}=M(u) B\, \\
\dot p&=&-\frac{\partial H}{\partial u}=f(u,A,B)\, \\
\dot B&=&-\frac{\partial H}{\partial A}=-M(u)A\, ,
\label{eq:ODE}
\end{eqnarray}
where the first two equations correspond to the generalized coordinates and the other two to the generalized momenta of the Hamiltonian system. The matrix $M(u)$ and the function for forces $f$ can be obtained trivially from Eqs.~\ref{eq:unddotReAnBn}--\ref{eq:AnBndot}. This structure allows the use of a splitting method\,\cite{hairerbook2006}, considering the two systems of ODEs:
\begin{eqnarray}
\mathrm{System\,\,I}\quad\left\{
\begin{array}{rcl}
\dot u&=&p\\
\dot A&=&0\\
\dot p&=&f(u,A,B)\\
\dot B&=&0
\end{array}\right.
\quad
\mathrm{System\,\,II}\quad\,\left\{
\begin{array}{rcl}
\dot u&=&0\\
\dot A&=&M(u)B\\
\dot p&=&0\\
\dot B&=&-M(u)A
\end{array}\right.
\label{eq:splitsystem}
\end{eqnarray}

System I corresponds to an ODE for the lattice variables while the charge variables are kept constant, and system II corresponds to an ODE for the charge variable while the lattice variables are kept constant. System I consists of a separable Hamiltonian $H=H(p)+H(q)$, and can be written as $\ddot u=f(u,A,B)$, which allows it to be integrated with the explicit Verlet method of second-order, also known as St\"ormer or leap-frog method.
%which is also symplectic.
It is given by:
\begin{eqnarray} %{rcl}
u^{n+1/2}&=& u^n+\frac{h}{2} p_n \\
p^{n+1}&=&p^n+h f(u^{n+1/2},A^n,B^n)\\
u^{n+1}&=&u^{n+1/2}+\frac{h}{2}p^{n+1}\\
A^{n+1}&=& A^n\\
B^{n+1}&=& B^n\, ,
\label{eq:LattIteration}
\end{eqnarray}
where $h$ is the step in time and $u^{n+1/2}$ is an intermediate point to calculate. We denote its flow by $\Phi_V^h$. As the charge variables do not change, the charge probability is trivially conserved.

System II is also  a separable Hamiltonian for the charge, i.e., $H_q= H(B)+H(A)$.
It is also linear and can be solved using the implicit midpoint rule given by:
\begin{eqnarray} %{rcl}
u^{n+1}&=& u^n\\
p^{n+1}&=& p^n\\
A^{n+1}&=& A^n+h M\big(\frac{u^{n+1}+u^n}{2}\big)\big(\frac{B^{n+1}+B^n}{2}\big)\\
B^{n+1}&=& B^n-h M\big(\frac{u^{n+1}+u^n}{2}\big)\big(\frac{A^{n+1}+A^n}{2}\big)\,.
\label{eq:ChargeIteration}
\end{eqnarray}
The values $A^{n+1}$ and $B^{n+1}$ are obtained by solving the second pair of linear equations above.

The lattice Hamiltonian is conserved in System II because $(u,p)$ do not change, so we only have to consider the part corresponding to the charge. The charge Hamiltonian and probability are given by quadratic expressions $H_q=\frac{1}{2}A^T M A+\frac{1}{2}B^T M B$ and  $|c|^2=A^T I A+B^T I B$, where $^T$ represents the transpose and $I$ the identity matrix. Both $M$ and $I$ (trivially) are symmetric. The implicit midpoint method conserves quadratic functions and, therefore,  the Hamiltonian and the charge probability.

Therefore, the charge probability is exactly conserved in both steps I and II and the Hamiltonian is exactly conserved in step II and in step I up to the second order in the integration step $h$.

The Strang or Marchus splitting\,\cite{hairerbook2006} obtains the global map in a symmetric way as:
\begin{eqnarray} %{rcl}
\Phi^h=\Phi^{h/2}_V \circ\Phi^{h}_{IM}\circ\Phi^{h/2}_{V}\, .
\label{eq:splitmethod}
\end{eqnarray}
The numerical map is time-reversible, conserves exactly the charge probability and the Hamiltonian to second order
in the integration step.
The global map conserves exactly the charge probability and the Hamiltonian to second order in the integration step.

Note that, if convenient, the splitting can be done by composing a larger number of flows. For example, it is possible to divide the IM step into as many IM steps as convenient. The same can be done with the V steps.
There are other splitting methods that could be used. For example, the splitting into five different steps is developed and applied to a related system in Refs.\,\cite{bajars-archilla2022A,archilla-bajars2023}.

\end{document}